\documentclass[acmsmall,screen,nonacm]{acmart}

\usepackage{microtype}
\usepackage{graphicx}
\usepackage{multirow}
\usepackage{mathtools}
\usepackage{amsthm}
\usepackage{stmaryrd}
\usepackage{mathpartir}
\usepackage{soul}
\usepackage[dvipsnames]{xcolor}
\usepackage[english]{babel}
\usepackage{algorithm}
\usepackage{algpseudocode}
\usepackage{cleveref}

\theoremstyle{plain}
\newtheorem{theorem}{Theorem}[section]

\newtheorem{lemma}[theorem]{Lemma}
\newtheorem{corollary}[theorem]{Corollary}
\theoremstyle{definition}
\newtheorem{definition}[theorem]{Definition}

\theoremstyle{remark}

\newcommand{\tali}{\textsc{Tali }}

\newcommand{\R}{\mathbb{R}}

\newcommand{\calP}{\mathcal{P}}
\newcommand{\calE}{\mathcal{E}}
\newcommand{\calV}{\mathcal{V}}
\newcommand{\calD}{\mathcal{D}}
\newcommand{\calC}{\mathcal{C}}
\newcommand{\calT}{\mathcal{T}}

\newcommand{\True}{\textsf{true}}
\newcommand{\False}{\textsf{false}}

\newcommand{\N}{\mathbb{N}}
\newcommand{\Z}{\mathbb{Z}}
\newcommand{\partto}{\rightharpoonup}

\newcommand{\judge}[3][\cdot]{#1 \vdash #2 : #3}

\newcommand{\dom}{\text{dom}}

\newcommand{\sig}[1]{\mathsf{sig}(#1)}

\newcommand{\dbrack}[1]{\llbracket #1 \rrbracket}
\newcommand{\muDenot}[2][{\rho,\phi}]{\mu\dbrack{#2}_{#1}}
\newcommand{\later}{\triangleright}
\newcommand{\upclos}[1]{\uparrow #1}
\newcommand{\bdepth}[1]{\mathrm{dp}(#1)}   %
\newcommand{\hgt}[1]{\mathrm{h}(#1)}        %
\newcommand{\satL}[1]{\models_{#1}}         %

\newcommand{\kw}[1]{\textcolor{Blue}{\mathsf{#1}}\;}
\newcommand{\fn}[1]{\textcolor{Brown}{\mathrm{#1}}}
\newcommand{\con}[1]{\textcolor{Orchid}{\mathsf{#1}}}
\newcommand{\flip}{\kw{bern}}
\newcommand{\bern}{\flip}  %
\newcommand{\geom}{\kw{geom}}
\newcommand{\cmnt}[1]{\textcolor{Gray}{\quad\leftarrow\textit{#1}}}
\newcommand{\letin}[3]{\kw{let} #1 = #2\;\kw{in} #3}
\newcommand{\ite}[3]{\kw{if} #1\;\kw{then} #2\;\kw{else} #3}
\newcommand{\defin}{\kw{define}}
\newcommand{\case}[4]{\kw{case} #1\; \kw{of} \overline{\{#2 (\overline{#3}) \to #4\}}}

\renewcommand{\sig}{\text{sig}}

\title{Stochastic Summaries: Characterizing Structured Datasets Using Probabilistic Programs}

\author{Jack Stanek}
\affiliation{%
  \institution{University of Wisconsin -- Madison}
  \city{Madison, WI}
  \country{USA}
}
\email{jstanek@cs.wisc.edu}

\author{Adithya Murali}
\affiliation{%
  \institution{University of Wisconsin -- Madison}
  \city{Madison, WI}
  \country{USA}
}
\email{adithyamurali@cs.wisc.edu}


\begin{document}

\begin{abstract}
What does my data look like? We study the problem of representing and learning \emph{summaries} of datasets that account for both their structural and statistical properties. This problem is widely useful across many application domains that involve large datasets and analysis/modeling of data, and is of particular interest to Programming Languages. In particular, in this work we focus on the problem of learning these ``stochastic summaries'' for tree-like data composed of inductive datatypes and 
argue that probabilistic programs over inductive datatypes can effectively represent the desired summaries. We present \textsc{Tali}, a Probabilistic Programming Language over inductive datatypes and study the problem of learning such programs from data samples. We identify a semantic property, called \emph{Observation-Bounded Compilability}, which identifies a set of recursive probabilistic programs which admit exact inference, and can represent many useful and complex probabilistic structures. We then propose an algorithm for learning \textsc{Tali} programs satisfying the OBC property from observations using a combination of program synthesis (stochastic search) and maximum likelihood inference. We implement our learning framework, evaluate it on a diverse benchmark suite, and record our observations on the merits of our approach. 
We demonstrate the utility of learning stochastic summaries as symbolically-represented priors for downstream learning tasks. We also showcase another utility by using our technique to learn operational profiles of real-world datasets.

\end{abstract}

\maketitle
\section{Introduction}
\begin{center}
    \emph{What does my data look like?}
\end{center}

Data distributions are a ubiquitous construct in computer science.
The most well known example of this is in machine learning, where the formal objective of learning problems is to obtain the natural ``data generating distribution''.

The representation, modeling, and analysis of large data sets is itself an important research question that powers several applications~\cite{saadBayesianSynthesisProbabilistic2019}. One view of the goal of such efforts in modeling data distributions is to construct \emph{summaries} or abstractions for them. Empirical datasets can be thought of as samples from an underlying ``true'' generative model.
In contrast to the more commonly studied logical or language-based abstraction mechanisms which capture structural properties or invariants, data distributions capture both statistical and logical/relational properties of the domain of interest.
These distributional abstractions can also be compositional: probabilistic graphical models, causal models, neural networks, and probabilistic programs are all examples of compositional languages for representing data distributions.

\paragraph{\bfseries Summaries for Data Distributions and Programming Languages} The representation and learning of probabilistic data summaries is also of specific interest to the field of Programming Languages. The design of abstraction spaces that admit expressive and compositional representations of practically useful data distributions while simultaneously being amenable to efficient learning is a goal of many research disciplines in PL, including program synthesis, probabilistic programming, and neurosymbolic programming/learning. 
This point has also been argued in several prior works in probabilistic programming, and in particular the work in~\cite{saadBayesianSynthesisProbabilistic2019} also pursues learning interpretable and operationalizable data summaries for time series data and tabular data.

Additionally, with stochasticity being increasingly woven into the vocabulary of modern system building and algorithm design~\cite{barbalho2023virtual,kraskaCaseLearnedIndex2018b,dong2020learning}, the study of stochastic data summaries can prove to be a useful tool. For example, prior literature has explored the concept of \emph{operational profiling} \citep{musaOperationalProfilesSoftwarereliability1993}, where one specifies a generative function that is meant to capture a program's usage patterns. Testing or analyzing systems against operational profiles (as opposed to the typical test suites obtained from analysis of the program's logic) can provide a useful perspective on the practical functionality and robustness of software~\cite{aroraSoftwareReliabilityImprovement2005}. %

\paragraph{\bfseries Synthesis of Probabilistic Programs over Structured Data}
We make two choices towards defining a formal problem.
First, we consider distributions over structured data, specifically generalized labeled trees in the form of inductive datatypes.
This choice is motivated by several factors.
First, computing with inductive datatypes is supported by many algorithms, and more generally, labeled trees are a ubiquitous construct in software.
Second, the structure of data can mirror information about its latent generative process, such as recursive patterns or local dependencies.
By contrast, consider integer programs, which can perform arithmetic operations that collapse the individual values (and therefore their distributions) into aggregates; intermediate distributions must be marginalized, which may be intractible or require approximate methods.
Finally, inductive datatypes allow us to study complex recursion patterns in generative processes that exploit repeated substructure.
Inductive datatypes are in fact an under-explored area in the study of distributions over structured data, with contemporary literature focusing mostly on tabular data \citep{papenbrockFunctionalDependencyDiscovery2015,xuModelingTabularData2019,apellanizImproved2024,tazwarTabVAENovelVAE2024}.

The second choice we make is to represent distributions as probabilistic programs.
This choice is also motivated by several considerations.
First, probabilistic programs are a natural choice for representing generative processes.
This perspective is itself not new;
in fact, it is one of the central ideas behind probabilistic programming at large \citep{goodmanChurchLanguageGenerative2014,gordonProbabilisticProgramming2014,meentIntroductionProbabilisticProgramming2021}, and more specifically has also been heavily utilized by the literature on testing \citep{tjoaTuningRandomGenerators2025,goldsteinParsingRandomness2022,goldsteinReflectingRandomGeneration2023,claessenQuickCheckLightweightTool2000,paveseInputs2018}.
Second, we are inspired by literature on operational profiling \citep{musaOperationalProfilesSoftwarereliability1993,hartmannStatisticalAnalysisOperational2016}, where the skeleton or sketch of the distribution is determined at least in part from knowledge of the data domain.
For example, a distribution over binary trees could utilize ``knobs'' relating the key stored at a node to whether it is a leaf, its height, or perhaps the keys stored at the children of the node.
This framing also allows us to utilize paradigms from the literature on program synthesis where learning is modeled as the search for an optimal candidate in a given grammar \citep{lakeHumanlevelConceptLearning2015,saadBayesianSynthesisProbabilistic2019,noriEfficientSynthesisProbabilistic2015}.
Third, as discussed above, programs can easily (and explicitly) represent complex recursion patterns over tree-like data compared to other choices of representation. Such programmatic representations can be easily integrated into existing workflows where data profiles are needed. For example, one can use a program learned from examples to generate realistic test cases.

\begin{quote}
\emph{In this paper we formalize stochastic summaries as a class of probabilistic programs over inductive datatypes and develop techniques for learning them from sample data. %
}
\end{quote}

\subsection*{Main Contributions}

We advance the above vision in this work via the following contributions:

\paragraph{\bfseries 1. Observation-Bounded Compilability}

The main contribution of this paper is the insight that productive recursion can be exploited to guarantee exact inference for recursive programs over inductive datatypes (\autoref{subsec:obsDepthAnalysis}).
This choice is motivated by our learning domain of labeled trees.
Productivity is the semantic property that every finite approximation of a possibly-infinite output (e.g., every finite prefix of a stream) is produced in finitely many steps \citep{atkeyProductive2013}.
In our setting, we wish to learn a program which fits a set of observations.
Conditionalizing a productive program's execution on the fact that a given output was generated, productivity allows us to ``walk'' the execution of the program in lock-step with structure of the observation, pruning dead execution paths which produce constructors that are not present in the observation.

We formalize this notion by defining the \emph{Observation-Bounded Compilability} property, which identifies a class of programs for which a given observation directly bounds the depth of recursive calls, and thus the number of proabilistic choices, which are consistent with the production of that observation.
Therefore, in contrast to statically bounded recursion, the OBC property picks out a \emph{parametrically bounded} class of programs for which each observation corresponds with a straightforward unrolling of the program.
This unrolling immediately enables both the exact inference algorithm noted in the previous paragraph, but also marginal likelihood techniques for program selection, as well as parameter tuning methods.
For values produced by an unambiguous program trace, the observation likelihood factors into a product over Bernoulli random variables, and we can quickly compute marginal likelihoods in closed form over beta priors \citep{gelmanBayesianDataAnalysis2013}.
For ambiguous observations, we can instead use approximate variational inference techniques \citep{bealVariationalBayesianEM2003,bleiVariationalInferenceReview2017} to compute lower bounds on the marginal likelihood.
Finally, given a program, we can construct a finite-size arithmetic circuit representing the likelihood function given the symbolic probabilistic parameters.
We can then use the circuit to tune parameters using maximum-likelihood estimation (MLE) via gradient-based methods.

\paragraph{\bfseries 2. Probabilistic Programs as Symbolic Priors for Inductive Data}

Our second contribution is the use of probabilistic programs as \emph{teacher distributions} for downstream learning tasks over inductive data.
Over the past decade, the machine learning field has seen an explosion in the use of large, scalable models such as transformers \citep{vaswaniAttention2017}.
By virtue of their capacity and generality, these models can learn very complex patterns, but this generality is a double-edged sword: they may need extraordinary amounts of data to generalize to unseen samples.
Probabilistic programs, on the other hand, directly denote an exact distribution with particular properties and constraints.
In settings where values have precisely defined logical properties or internal invariants, recursive probabilistic programs are a natural way to construct generative models.

To this end we develop a program synthesis technique which fits recursive probabilistic programs as generative models of real-world data.
Our synthesis algorithm enumerates recursive program structures which support the training set, using parent annotation \citep{johnsonPCFGModelsLinguistic1998} and Bayesian model merging \citep{hwangInducingProbabilisticPrograms2011,stolckeInducingProbabilisticGrammars1994} to identify local data dependencies;
enumerated proposals are scored by marginal likelihood.
Our method then uses a stochastic Metropolis-style search over the program space using subtree replacement to locate additional candidate programs, which are scored in similar fashion.
During synthesis, we record the top-$k$ candidate programs, and at the end of the process we fit parameters by maximum likelihood using gradient-based methods, returning the best-fit program.

We evaluate the effectiveness of our synthesis algorithm on real-world datasets from the software engineering literature \citep{davisWhyArentRegular2019,malyshevGettingMostOut2018}.
We execute our synthesis algorithm on these data, and evaluate the effectiveness of using synthesized programs as \emph{symbolic priors} over the data.
Our method uses the synthesized program as a generator of large training corpora;
we are thus using the program synthesis step as a way to identify and amplify value constraints in the data, where a pretrained transformer with an uninformative pretraining prior would require a much larger training set to identify.
We focus on the low-data regime, evaluating the method on 10-300 training samples.
We find that pretraining on a synthetic corpus produces a model which fits the data better than the the program itself on average.
Our method is also able to identify local and contextual invariants within structured data, such as ordering constraints.

\paragraph{\bfseries 3. The \tali Probabilistic Programming Language}
Our final contribution is the implementation of \tali, a simple first-order probabilistic programming language over inductive datatypes.
We give a formal account of the syntax and semantics of our language in~\autoref{sec:dslAndTypeSystem}, and describe a type system, adapted from \citet{atkeyProductive2013}, which guarantees that well-typed programs exhibit the OBC property.
\tali supports unbounded recursion, inductive datatypes, standard control flow constructs, and exact inference over observed samples.
We also implement an copmiler which constructs arithmetic circuits representing observations' likelihood functions over symbolic parameters.

\section{Overview}
\label{sec:overview}

This work centers on the following high-level question: given a dataset of structured, inductively defined values, what would a compact and interpretable summary of that dataset look like?
Consider the probabilistic program in \autoref{fig:sexpSampler}, which models a generative process that produces S-expressions according to a simple probabilistic grammar.
It generates atom constructors both stochastically and from a rotating deterministic set of atoms.
Each flip in the function body contributes to some aspect of the return value's structure or content: for instance, the results of $\flip \theta_1$ and $\flip \theta_4$ determine the type of S-expression that is generated, while $\flip \theta_2$ and $\flip \theta_3$ determine which atom is produced in the $\con{SAtom}$ branch.
For example, if $\fn{next}(\con{AtomC}) = \con{AtomA}$, suppose we evaluate the call $\fn{sampleSexpr}(\con{AtomC})$ with probabilistic choices $\flip \theta_1 = \False$, $\flip \theta_4 = \False$, $\flip \theta_2 = \True$, and finally $\flip \theta_1 = \True$ on the recursive call.
Such an execution would return:

\centerline{$\con{Cons}(\con{SAtom}(\con{AtomA}), \con{SAtom}(\con{AtomA}))$}

Despite its brevity, our stochastic sampler represents a nontrivial probability distribution with infinite support while remaining highly interpretable: its meaning can be understood simply by reading the program.
Neural models over serialized trees could fit the same distribution, but their parameter weights would offer little insight into its recursive structure.
The probabilistic program, by contrast, explicates this structure directly.

\begin{figure}[t]
    \centering
    \[
    \begin{array}{l}
        \defin \fn{sampleSexpr} (a: \con{Atom}) : \con{Sexpr} = \\
        \quad \ite{\flip \theta_1}{\\%
        \qquad \con{SAtom} (\ite{\flip \theta_2}{a}{\\%
        \qquad \quad \ite{\flip \theta_3}{\con{AtomA}}{\con{AtomB}}})\\\quad}{\ite{\flip \theta_4}{\\%
        \qquad \con{Nil}}{\\%
        \qquad \con{Cons}(\fn{sampleSexpr}(\fn{next}(a)), \fn{sampleSexpr}(\fn{next}(a))})}\\
    \end{array}
    \]
    \caption{Stochastic S-expression sampler. Atoms are sampled both at random and from a deterministic set of rotating atom choices. Each probabilistic choice contributes to some aspect of the return value's structure and content.}
    \label{fig:sexpSampler}
\end{figure}

\paragraph{\bfseries Probabilistic Programs as Hypothesis Spaces} The structure of data can often encode information about the underlying generative process. %
For example, in our S-expression generator, the program constructs an observation step-by-step, with each recursive call and probabilistic choice corresponding to a particular node in the expression tree.
More complex tree structures may have more sophisticated internal constraints than this S-expression generator, but the principle remains the same: structure is a direct result of the probabilistic choices made by the program.
Indeed, existing neural program learning systems such as \textsc{DreamCoder}~\citep{ellisDreamCoderBootstrappingInductive2021} and earlier Bayesian learning techniques~\citep{lakeHumanlevelConceptLearning2015} frame the learning task explicitly as a structure discovery problem over data.

The space of recursive probabilistic programs is expressive, but it also includes programs with undesirable properties, such as nontermination or wasteful computation.
For instance, our S-expression sampler could have been na\"ively written to generate unrestricted random S-expressions, iterating until a particular condition is satisfied---for example, until certain ground atoms appear in a particular order.
Such programming patterns can make probabilistic inference over observations difficult or impossible.
In general, efficient inference is hampered by path explosion, marginal distributions without analytic closed forms, or undecidability.
In order to avoid such pathological programs, a well-behaved subclass of programs is needed.

\paragraph{\bfseries Efficiently Learnable Data Generators} %
The key observation we leverage in this paper is that, for certain well-behaved programs, sources of randomness can be mapped cleanly to aspects of the sampled values. %
Returning to our S-expression example, consider computing the likelihood of observing $\con{Cons}(\con{SAtom}(\con{AtomA}), \con{SAtom}(\con{AtomA}))$.
The top-level call consumes two flips, $\flip \theta_1$ and $\flip \theta_4$, resolving the $\con{Cons}$ branch.
Each recursive sub-call unrolls one level further, consuming $\flip \theta_1$ and $\flip \theta_2$ to resolve the $\con{AtomA}$ branches.
The depth of unrolling is determined entirely by the structure of the observation itself: a value of depth two requires two levels of unrolling (though the lower level of sub-calls contains two independent calls).
This is loosely analogous to backpropagation-through-time (BPTT) in the context of recurrent neural networks (RNNs)~\citep{werbosBackpropagationTimeWhat1990}, where a bounded unrolling approximates inference by truncating recursion depth, trading exactness for tractability.
Unlike BPTT, where truncation introduces approximation error, well-behaved programs need not sacrifice exactness: because the space of random choices is bounded by the observation, a truncated computation can remain exact. We formalize this insight as a property called \emph{Observation-Bounded Compilability} in~\autoref{subsec:sublangAndOBC}.

\section{Background and Problem Definition}
\label{sec:problem}

In this paper we use the convention that syntactic elements annotated with an overline (for instance, $\overline{e}$) refer to finite sequences of those elements.
In inference rules, this notation indicates that the rule, premise, or conclusion is implicitly quantified over the elements of the sequence.
A superscripted overline (e.g. $\overline{e}^n$) indicates that there are exactly $n$ elements in the sequence.

\subsection{Semantic domain}
Our semantic domain is a set of finite labeled trees, which we take to be the \emph{observable values}, meaning the objects that a dataset consists of, and consequently the objects against which our semantics assigns likelihoods.
(The semantics of \autoref{subsec:probSemantics} works in a slightly larger set, generalizing the observable values to partially-observed values containing placeholders for parts of a value that have not been produced yet.
The observable values reappear there as the maximal elements of this set.)

We fix a finite set of sorts $\calT$ and a countable ranked alphabet of finitary constructor symbols $\calC$, forming a multi-sorted signature $(\calT, \calC)$.
We refer to the sorts of $\calT$ as ``types,'' and each constructor is associated with a unique type.
We denote the set of constructors associated with type $T$ as $\calC_T$, and require that each type has at least one associated constructor; that is,
\[
    \biguplus_{T \in \calT} \calC_T = \calC.
\]
The domain of values of type $T \in \calT$ is the set of well-sorted finite ground terms of type $T$ over $\calC$.
We write $C(\overline{v})$ to denote a value with constructor $C$ and subterms $\overline{v}$ matching the signature of $C$, and use the notation $|C|$ to mean the arity of $C$.
Since every constructor has finite arity, every value is finitely branching, even though $\calC$ itself may be countably infinite.
We assume two distinguished types: \textsf{Bool}, with two nullary constructors, \True\,and \False, and \textsf{Int}, with one nullary constructor for each integer.
We denote the full set of all semantic values as $\calV$.
We define the function $\sig : \calC \to T^*$ to map constructors to their signatures.

\subsection{Sketch-based probabilistic program synthesis}

Let $\calP$ be a domain-specific probabilistic programming language, and denote $\calE \subseteq \calP$ as the set of expressions in $\calP$.
The language $\calP$ includes \emph{partial programs}, that is, programs with holes; a \emph{program sketch} in $\calP$ may have one or more holes.
Programs may also have \emph{symbolic parameters}, which are unknown at compile time.
For a program sketch $P \in \calP$ with $h$ holes and $k$ symbolic parameters, we write $P[\overline{e}|\theta]$ to mean the complete program obtained by filling the holes in $P$ with a vector of subexpressions $\overline{e} \in \calE^h$ and binding the symbolic parameters to the parameter vector $\theta \in \R^k$.
Let $\dbrack{\cdot} : \calP \to \calV \to [0, 1]$ be a semantics for $\calP$ that maps each complete program $P^\star$ (without unfilled holes or unbound symbolic parameters) to a sub-probability distribution over semantic values in $\calV$.
Let $D = \{v_i\}_{i=1}^{n}$ be a multiset of observations such that each $v_i \in \calV$.

The \emph{sketch-based probabilistic program synthesis} problem is to find, given a program sketch $p$ with $h$ holes and $k$ symbolic parameters, a vector of subexpressions $\overline{e}^\star \in \calE^h$ and a parameter vector $\theta \in \mathbb{R}^k$ that maximize the likelihood of the observations in $D$:
\[
    \langle \overline{e}^\star, \theta \rangle = \arg\max_{\overline{e} \in \calE^h, \theta \in \R^k} \prod_{i=1}^{n} \dbrack{P[\overline{e}|\theta]}(x_i)
\]

Equivalently, the objective can be written as the joint conditional probability of the observed data given a complete parameterized program:

\[
    \langle \overline{e}^\star, \theta \rangle = \arg\max_{\overline{e} \in \calE^h, \theta \in \R^k} \Pr \left( D \mid p[\overline{e}|\theta]\right)
\]

\section{The \tali Language}
\label{sec:dslAndTypeSystem}

In this section we introduce \tali, a probabilistic programming language over inductive datatypes for expressing probabilistic data summaries.
We first present the syntax and semantics of our language in~\autoref{subsec:syntax} and~\autoref{subsec:probSemantics}, respectively.
We then describe, in~\autoref{subsec:sublangAndOBC} and \autoref{subsec:obsDepthAnalysis}, a type system that defines a class of programs amenable to efficient learning: their likelihoods can be computed directly without marginalizing over infinite sets of execution traces.

\subsection{Syntax}
\label{subsec:syntax}

\begin{figure}[t]
    \centering
    \begin{align*}
        \text{Base Types:} \quad A &::= \mathsf{Bool} \mid \mathsf{Int} \mid T \\
        \text{Modal Types:} \quad \tau &::= \later^n A \qquad (n \in \N,\;\later^0 A \triangleq A)\\
        \text{Function Types:} \quad \sigma &::= \overline{\tau} \to \tau' \\
        \text{Parameters:} \quad p &::= \theta \mid \pi\\
        \text{Expressions:} \quad e &::= x \mid f(\overline{e}) \mid \letin{x}{e_1}{e_2}\\
        &\mid C \mid C(\overline{e}) \mid \case{e}{C}{x}{e'} \\
        &\mid \flip p \mid \geom p\\
        \text{Values:} \quad v &::= C \mid C(\overline{v}) \\
        \text{Definitions:} \quad d &::= \defin f (\overline{x: \tau}) : \tau' \coloneq e \\
        \text{Programs:} \quad P &::= e \mid d;\,P\\
    \end{align*}
    \caption{Syntax of the core calculus for recursive data sampling.}
    \label{fig:coreSyntax}
\end{figure}

\autoref{fig:coreSyntax} shows the syntax.
\tali is a first-order probabilistic programming language with recursive functions, pattern matching, standard control-flow constructs, variable assignment, and probabilistic choice operators.
A program is a sequence of function definitions followed by an expression which represents the main body of the program.
Stochasticity is introduced through the probabilistic choice operators $\flip p$ and $\geom p$.
The $\flip p$ expression samples a Bernoulli distribution with parameter $p$, evaluating to $\textsf{true}$ with probability $p$ and $\textsf{false}$ with probability $1 - p$.
Similarly, the $\geom p$ expression samples a geometric distribution with parameter $p$, evaluating to a (zero-indexed) natural number $n$ with probability $p^{n}(1 - p)$.

Parameters are not first-class values and are fixed at runtime.
A given parameter may either be a concrete numerical value $\theta$ on the interval $(0, 1)$, or else a symbolic parameter $\pi$.
Symbolic parameters are mapped to numerical values just like concrete parameters, but their values are supplied by the environment at runtime.

Unlike many other probabilistic languages, we do not support an \textsf{observe} construct.
This is primarily because we are specifically interested in learning generator programs from observations.
In this application domain, the main use of observe statements would be to capture logical properties of the data domain (e.g., that the trees are valid binary search trees).
Synthesizing logical properties from observations is challenging in its own right, as is designing a language that incorporates both logical and distributional properties (and \textsf{observe} statements may not be the appropriate solution). We leave this problem to future work.
Note that one can ensure logical properties in a constructive manner using program patterns that only generate satisfying data values (e.g., always increment to ensure sortedness), and our framework supports this modeling.

Types are stratified into \emph{base} types, which are the ordinary types of values, and \emph{modal} types $\later^n A$, which additionally record that a value is available only $n$ steps ``later.''
We defer full explanation of the later modality to \Cref{subsec:obsDepthAnalysis}.

\subsection{Semantics}
\label{subsec:probSemantics}

A \tali program denotes a sub-probability distribution over values that it can generate.
Because programs are recursive and their outputs are inductive, we must account for programs that may diverge or produce infinite values.
\label{ex:gen}
Consider, for example, the program
\[
  \defin \fn{gen}() : \text{List} := \con{Cons}(0, \fn{gen}());\;\fn{gen}()
\]
This program clearly does not terminate, and it should assign zero likelihood to every finite list, since all of its probability mass rests on infinite lists.
Because our semantic domain is restricted to the finite values in the term algebra induced by the set of constructors, the na\"ive semantics over complete values are ill-suited to handle all \tali programs.
To formalize the semantics of \tali, we instead interpret a program as a distribution over \emph{partially evaluated values}, whose as-yet-unproduced parts are represented by a placeholder $\bot$.
The (sub-)probability distribution over complete values (and thus over observed values) is then the special case over $\bot$-free values.

\begin{definition}[Partial value]
    A \emph{partial value} is a value in which some subterms may be replaced with $\bot$, meaning ``not yet evaluated.''
    \[
        \delta \Coloneqq \bot \mid C(\overline{\delta})
    \]
\end{definition}

We define an ordering on partial values $\delta_1 \sqsubseteq \delta_2$, read ``$\delta_1$ is no more defined than $\delta_2$,'' such that $\bot \sqsubseteq \delta$ for all $\delta$, and $\delta_1 \sqsubseteq \delta_2$ if $\delta_1 = C(\overline{\delta_{1i}})$ and $\delta_2 = C(\overline{\delta_{2i}})$ where $\delta_{1i} \sqsubseteq \delta_{2i}$ pointwise.
Values with distinct constructors are incomparable.
We say $\delta_1$ and $\delta_2$ are \emph{compatible} if they share an upper bound; intuitively, this means that they agree on their top-level constructors and all transitive subterms, up to $\bot$.
We define $\calV_\bot$ to mean the set of all partial values generated by the fixed signature as described in \autoref{sec:problem}.
The members of $\calV_\bot$ are finite, so $\calV_\bot$, like $\calV$, is countable.
The set of semantic values $\calV \subseteq \calV_\bot$ is then the special case of $\bot$-free (and thus maximal) partial values.
We refer to such values $v \in \calV$ as \emph{complete values}.

\subsubsection{Probability distribution over partial values}
\begin{figure}
    \begin{mathpar}
        \muDenot{C(\overline{e})}(C(\overline{\delta})) =
            \prod_{i=1}^{|C|} \muDenot{e_i}(\delta_i)\and
        \muDenot{C(\overline{e})}(C'(\overline{\delta})) = 0,\, C \neq C'\and
        \muDenot{C(\overline{e})}(\bot) = 0\and
        \muDenot{x}(\delta) = \begin{cases}
            1&\rho(x) = \delta\\
            0&\text{otherwise}\\
        \end{cases}\and
        \muDenot{\bern p}(\delta) = \begin{cases}
            p&\delta = \True\\
            1-p&\delta = \False\\
            0&\text{otherwise}\\
        \end{cases}\and
        \muDenot{\geom p}(\delta) = \begin{cases}
            (1 - p)p^{\mathcal{N}\dbrack{\delta}}&\delta = 0, 1, 2, \dots\\
            0&\text{otherwise}
        \end{cases}\and
        \muDenot{f(\overline{e})}(\delta) =%
            \sum_{\overline{\delta'} \in \calV_\bot^{|f|}}%
                \left(\prod_{i=1}^{|f|} \muDenot{e_i}(\delta_i')\right) %
                \cdot \phi(f)(\overline{\delta'})(\delta)\and
        \muDenot{\letin{x}{e_1}{e_2}}(\delta) =%
            \sum_{\delta' \in \calV_\bot} \muDenot{e_1}(\delta')%
                \cdot \muDenot[{\rho[x \mapsto \delta'], \phi}]{e_2}(\delta)\and
        \muDenot{\case{e}{C_i}{x_i}{e_i'}^n}(\delta) =%
            \sum_{i=1}^n \sum_{\overline{\delta'} \in \calV_\bot^{|C_i|}}%
                \muDenot{e}(C_i(\overline{\delta'})) \cdot%
                \muDenot[{\rho[\overline{x} \mapsto \overline{\delta}],\phi}]{e_i'}(\delta)%
                + \begin{cases}
                    \muDenot{e}(\bot)& \delta = \bot\\
                    0& \text{otherwise}\\
                \end{cases}
    \end{mathpar}

    \caption{%
        Denotational semantics of the \tali language.
        Each expression denotes a measure $\mu_{\rho,\phi}\dbrack{e}$ on partial values $\delta \in \calV_\bot$.
        $\mathcal{N}\dbrack{\delta}$ is a partial function which maps syntactic integers to $\mathbb{Z}$.
        We write $|f|$ to mean the arity of a function $f$ (that is, $|f| = |\overline{\tau}|$ if $\Sigma(f) = \overline{\tau} \to \tau'$).
    }
    \label{fig:semantics}
\end{figure}

A program $P = \overline{d}; e$ denotes a discrete sub-probability distribution over $\calV_\bot$.
Its denotation gives the likelihood that execution of $P$ eventually produces a partial observation $\delta$.
To formally construct this denotation, we first define a measure on $\calV_\bot$ denoted by an expression $e$.
The equations are given in \autoref{fig:semantics}.
The measure $\muDenot{e}$ is indexed by a variable context $\rho : \text{Name} \partto \calV_\bot$ and a function context $\phi : \text{Name} \partto (\calV_\bot^* \to \calV_\bot \to [0, 1]).$
Note that the variable context $\rho$ binds variables to partial values, not to distributions over partial values.
Threading dependent state is handled through marginalization in the equations for the binding forms.
This distinction matters because it prevents incorrect denotations of expressions such as $\letin{x}{\flip p}{\con{Pair}(x, x)}$.
Were a na\"ive distribution-driven approach used, we would need to take care to ensure that the pair's members are correctly correlated and not independent Bernoulli trials.
The state-threading approach naturally sidesteps such pitfalls.

For a well-formed expression $e$ in which every $\kw{case}$ subexpression is exhaustive, the measure in \autoref{fig:semantics} is a probability distribution:

\begin{lemma}[Probability distribution]
    \label{lem:probDist}
    Let $e$ be a well-formed expression.
    Let $\rho$ be a variable context defined on the free variables of $e$, and let $\phi$ be a function context whose values are probability distributions.
    Then $\muDenot{e}$ is a probability distribution over $\calV_\bot$; that is, \[\sum_{\delta \in \calV_\bot} \muDenot{e}(\delta) = 1.\]
\end{lemma}

We can now define the likelihood that a program is \emph{compatible} with a particular finite, possibly partial observation.
For a partial value $\delta$, we interpret its upward closure $\upclos{\delta} = \{\delta' \in \calV_\bot \mid \delta \sqsubseteq \delta'\}$ as the set of partial values compatible with the prefix $\delta$; we call this set a \emph{demand}.
The denotation of an expression $e$ on a partial value $\delta$ is the measure of $\upclos{\delta}$.
We write $\dbrack{e}_{\rho,\phi}(\delta)$ for the likelihood that $e$ meets the demand $\delta$:
\[
    \dbrack{e}_{\rho,\phi}(\delta) = \muDenot{e}(\upclos{\delta}) =%
        \sum_{\delta_i \in \upclos{\delta}}\muDenot{e}(\delta_i)
\]
Since $\upclos{\bot} = \calV_\bot$ and $\muDenot{e}$ is a probability distribution on $\calV_\bot$, we have $\dbrack{e}_{\rho,\phi}(\bot) = 1$.
Furthermore, because every value in $\calV$ is maximal, the two denotations agree exactly on complete values:
indeed, if $\delta \in \calV$, then ${\upclos{\delta} = \{\delta\}}$, and so $\muDenot{e}(\delta) = \dbrack{e}_{\rho,\phi}(\delta)$.
Since $\muDenot{\cdot}$ is a probability distribution on $\calV_\bot$, $\dbrack{\cdot}_{\rho,\phi}$ is a sub-probability distribution on $\calV$.
We will focus on $\dbrack{\cdot}_{\rho,\phi}$ for the rest of this paper, as it is the value that we compute to evaluate the likelihood of a fully observed set of values for a given program.

\subsubsection{Function fixed points and approximants}
\label{subsec:fixedPoints}

Next, we define the semantics of recursive functions.
We handle recursion in the standard way, by taking a fixed point of a function context.
For a program $P$, written
\[
    P = \defin f_1(\overline{x_1 : \tau_1})  : \tau_1' \coloneq e_1;\;\dots\;%
        \defin f_m(\overline{x_m : \tau_m})  : \tau_m' \coloneq e_m;\;%
        e_{\text{main}},
\]
we define a functional whose value for each $f_i$, with parameters $x_{i1}, \dots, x_{ik_i}$ and $\overline{\delta} \in \calV_\bot^{k_i}$, is:
\[
    \Phi(\phi)(f_i)(\overline{\delta})(\delta') = \mu_{\rho_i,\phi}\dbrack{e_i}(\delta') \quad %
        \text{where } \rho_i = \{x_{ij} \mapsto \delta_j\}_{j = 1}^{k_i}
\]
We define $\phi_0(f)(\overline{\delta})$ as the point mass on $\bot$:
\[
    \phi_0(f)(\overline{\delta})(\delta) = \begin{cases}
        1&\delta = \bot\\
        0&\text{otherwise}
    \end{cases}
\]
and define $\phi_{k+1} = \Phi(\phi_k)$.
Operationally, the context $\phi_k$ represents iteratively ``unrolling'' each function call $k$ times, placing any residual mass on $\bot$ at the base case.
Each unrolled context is an approximant of the fixed point of $\Phi$; we refer to each $\phi_k$ as the $k$-th approximant of the function environment.

The approximants need not converge pointwise as distributions, but their likelihoods converge monotonically.
\begin{theorem}[Refinement]
    \label{thm:refinement}
    Let $k \leq k'$ and $\rho \sqsubseteq \rho'$ pointwise.
    Then for any expression $e$, there exists a coupling of
        $\mu_{\rho,\phi_k}\dbrack{e}$ and $\mu_{\rho',\phi_{k'}}\dbrack{e}$,
        supported on pairs $(\delta, \delta')$, such that $\delta \sqsubseteq \delta'$.
    Thus $\dbrack{e}_{\rho,\phi_k}(\delta) \leq \dbrack{e}_{\rho',\phi_{k'}}(\delta)$ for each demand $\delta$,
    and approximants $\dbrack{e}_{\rho,\phi_k}$ are non-decreasing in $k$.
\end{theorem}

A consequence of \autoref{thm:refinement} is that repeatedly unrolling function applications only refines the output; it never revises what is already defined.
Thus we may define
\[
    \dbrack{e}_\rho(\delta) = \lim_{k \to \infty} \dbrack{e}_{\rho,\phi_k}(\delta) = \sup_{k} \dbrack{e}_{\rho,\phi_k}(\delta)
\]
as simply the limit of a bounded monotone sequence of reals.
The semantics of a full program $P$ is then
\[
    \dbrack{P}(\delta) = \dbrack{e_{\text{main}}}_\varnothing(\delta),
\]
and we write the $k$-th approximant of the semantics (i.e. the semantics evaluated under function environment $\phi_k$) as
\[
    \dbrack{P}_k(\delta) = \dbrack{e_{\text{main}}}_{\varnothing,\phi_k}(\delta).
\]

The semantics of \tali are well-defined for all recursive programs and partial values, although not all programs generate complete, $\bot$-free values $v \in \calV$ with probability 1.
Probability mass can escape from $\calV$ in two ways.
First, a program may encode a supercritical branching process which diverges with positive probability, or equivalently places mass on infinite trees (which we represent as sets of ever-deeper partial values truncated at $\bot$ leaves).
Such programs may still produce finite values, to which our semantics correctly assigns likelihoods; these likelihoods simply do not sum to 1.
Second, a program may produce no output, looping forever before deciding its first constructor.
Such programs place some or all of their probability mass on $\bot$; they are pathological because computing a likelihood requires marginalizing over a potentially unbounded set of execution traces consistent with the observation.
In the following section, we define a semantic program property, \emph{observation-bounded compilability}, that rules out such programs, along with a type system that accepts only well-behaved programs.

\subsection{Observation-Bounded Compilability}
\label{subsec:sublangAndOBC}

Our language features unbounded recursion in order to concisely express recursive generative processes over inductive values.
However, unbounded recursion can be problematic and difficult to deal with.
To guarantee that bounded unrolling suffices to compute a recursive program's distribution, we must ensure that unobserved probabilistic choices are well-behaved and can be marginalized exactly.

It is therefore desirable to identify a class of probabilistic programs with this property.
One common approach to recursion in probabilistic languages is a fixed unroll or iteration bound.
While this approach can help simplify implementation, it can produce incorrect likelihoods if the program's approximated likelihood does not stabilize at some finite unroll bound.
Worse, the minimum unroll bound needed to compute a value's likelihood exactly may be difficult to derive.
For instance, with fixed unrolling, one could increase the bound until observation likelihoods stabilize.
However, this could require recomputing the same likelihoods many times and might not converge for all recursive programs.
(As an example, programs with non-terminating recursion will never see their observed likelihoods plateau at the exact probability mass, though they may approach it asymptotically.)
Thus, it is desirable to statically ensure that \emph{all} \tali programs admit exact inference using a simple unrolling scheme.

Formally, we define the following \emph{observation-bounded compilability property} for recursive programs.
It characterizes programs for which inference can be performed exactly by unrolling function calls up to a bound determined by the observation.

\begin{definition}[Observation-bounded compilability]
    \label{def:obsBoundComp}
    A program $P$ is \emph{observation-bounded compilable} if and only if there exists a total computable function $\beta_P : \calV \to \N$ such that
    \[
        \forall v \in \calV, k \geq \beta_P(v): \dbrack{P}_k(v) = \dbrack{P}(v).
    \]
\end{definition}

Intuitively, the sorts of programs which satisfy this property are programs for which each recursive call is accounted for in the output.
Each ``layer'' of an observed value can be thought of as the result of a bounded set of probabilistic choices, so the set of program traces consistent with an observation can be recovered by iteratively ``peeling off'' constructors and recording which choices were made at each step.
This structure aligns naturally with real-world recursive samplers.
It rules out certain program behaviors, such as infinite loops or filters over infinite streams or state spaces, that could harm performance.
By imposing this restriction, we apply an inductive bias which guides synthesis towards programs that provide algorithmic or constructive explanations in terms of generative processes, rather than declarative logical invariants.

One important note is that the observation-bounded compilability property does not make probabilistic inference tractable in general.
Discrete probabilistic inference over boolean domains is \#P-hard \cite{rothHardnessApproximateReasoning1996}, and OBC does not ameliorate this.
The OBC property merely provides a bound on the length of an execution trace for an arbitrary finite observation.

\subsection{Guarded recursion}
\label{subsec:obsDepthAnalysis}

Enforcing OBC requires a single semantic restriction: every recursive call must produce observable output.
Since the size of an observation is finite, it follows that its size bounds the number of recursive calls needed to produce it.
Fortunately, guarded recursion provides a simple way to enforce this semantic property: every recursive call must be enclosed by one or more constructor applications.

To see why guardedness is needed, consider the following function definition, which recursively constructs a list of boolean values.

\begin{gather*}
    \begin{array}{l}
        \defin\;\fn{buildList}\;() : \text{BoolList} = \\
        \quad \ite{\flip p}{\\%
        \qquad \quad \con{Cons}(\flip q, \fn{buildList}())}{\ite{\flip r}{\\%
        \qquad \quad \con{Nil}}{\\%
        \qquad \quad \fn{buildList}() \quad \cmnt{unguarded recursion}}}
    \end{array}
\end{gather*}

The final recursive call to $\fn{buildList}$ is \emph{unguarded} because it is returned directly, without a constructor.
Suppose we wish to compute the likelihood of a given list, say $\con{Cons}(\True, \con{Nil})$.
There are many unrollings consistent with the observation; for instance, it is possible that $\flip p$, $\flip q$, and $\flip r$ all produced $\True$, resulting in the observation.
However, $\flip r$ could also produce $\False$ any number of times before producing $\True$. Thus, infinitely many traces are consistent with the observation, and no finite unrolling can compute its likelihood exactly.

We formalize this principle with a modal type system; selected rules are shown in \autoref{fig:typeRules}.
Since functions in \tali are first-order, programs are typed against a function signature $\Sigma$, which is a finite map from (unique) function names to type signatures.
Functions are added to $\Sigma$ in order of definition; mutual recursion is not directly supported.\footnote{Note that mutual recursion can be emulated using single recursion and a tag argument.}
We use the top-level judgment $\judge[\Sigma]{P}{\tau}$ to state a program's type: under function signature $\Sigma$, program $P$ is well-typed and produces a value of type $\tau$.
Expressions are typed with the judgment
\[
    \judge[\Sigma; \Gamma; f]{e}{\tau},
\]
meaning that under function signature $\Sigma$, variable context $\Gamma$, and enclosing function name $f$, the expression $e$ has type $\tau$.
We write $f = \cdot$ when $e$ is the main body of a program rather than a function body.

We now implement guarded recursion using the modal features of \tali's type system.

\subsubsection{Later modality}
\label{subsec:guardedRecursion}
To enforce guardedness, \tali's type system includes a clocked ``later'' modality, written $\later^n A$ .
Our approach is closely related to those of \citet{nakanomodality2000a} and \citet{atkeyProductive2013}, where similar systems are used to prove the productivity of corecursive computations.
Intuitively, a value at later index $n$ is one that will only be available after $n$ further steps; a value at index 0 is available ``now.''
Types are stratified; that is, $\later^n$ is a prefix attached to a base type rather than a type former in itself.
Consequently, there are no nested modalities (e.g. $\later^m (\later^n A)$), though modalities may be shifted (e.g. $\later^{m+n} A$).
Furthermore, since we take $\later^0 A$ to be literally the base type $A$, the types $A$ and $\later^0 A$ may be exchanged freely.

The guardedness property now follows from a short argument.
The index is raised by \textsc{T-Call} in recursive contexts, by \textsc{T-Case} in case destructuring, and by \textsc{T-Later-Sub} as a subsumption mechanism; it is lowered only by \textsc{T-Constr}.
Since the modal index of the return value of a recursive call is raised by at least 1, an unguarded recursive call cannot be returned directly since a modal type is never subsumed by a lower modal index, only by a higher one.
Furthermore, a recursive self-call cannot be ``laundered'' by wrapping it in a constructor and immediately destructuring it with a case expression; doing so leaves the modal index at its incremented value.
Thus every self-call occurring in a well-typed body must be enclosed in at least one constructor application before being returned, which is precisely what the $\fn{buildList}$ example above violates.
Its final $\fn{buildList}()$ call is returned directly without being wrapped in a $\con{Cons}$, and this fails to type-check since the call is typed at index 1 in a position demanding index 0.

\subsubsection{Typing judgments}
The selected rules of \autoref{fig:typeRules} show how the modal index is manipulated.
\textsc{T-Later-Sub} defines the subsumption discipline: types may always be used at a later index (though not at an earlier one).
The modal index is incremented by function calls and case destructuring.
\textsc{T-Call} types recursive calls at a higher modal index, thereby incurring some amount of ``modal debt.''
(Non-recursive calls are typed at the same index, merely passing the index of their arguments through to its result.)
\textsc{T-Case} binds the children of a scrutinee at modal index $n$ to variables at modal index $n + 1$.
This is necessary since the scrutinee's children may not be available yet, despite the scrutinee's root constructor being available;
in effect, performing case analysis defers the modal obligation by one step.
\textsc{T-Constr} is the sole elimination rule for modal types: a constructor applied to arguments at index $n+1$ yields a result at index $n$, discharging one unit of modal debt in exchange for one layer of observable output.

\subsubsection{Type soundness}

We now state the main soundness property of the type system: well-typed programs exhibit the OBC property.

\begin{theorem}[Soundness]
    \label{thm:correctness}
    For any program $P$ and type $\tau$, if $\judge[\cdot]{P}{\tau}$, then $P$ is OBC.
\end{theorem}

A key consequence of OBC is that probabilistic inference for such programs can be implemented by unrolling function definitions to a depth determined by the observation.
In the following section, we briefly describe a straightforward inference procedure for well-typed programs.

\begin{figure}[t]
    \begin{mathpar}
        \inferrule[T-Later-Sub]{
            \judge[\Sigma; \Gamma; f]{e}{\later^m A}\\
            m \leq n
        }{
            \judge[\Sigma; \Gamma; f]{e}{\later^n A}
        }\and
        \inferrule[T-Constr]{
            C \in \calC_A \\
            \sig(C) = (A_1, \dots, A_k)\\
            \forall i \leq k.\; \judge[\Sigma; \Gamma; f]{e_i}{\later^{n+1} A_i}
        }{
            \judge[\Sigma; \Gamma; f]{C(\overline{e})}{\later^{n} A}
        }\and
       \inferrule[T-Call]{
            j \geq 1 \text{ if } g = f \text{ else } j \geq 0\\
            \Sigma(g) = (\later^{m_1} A_1, \dots, \later^{m_k} A_k) \to \later^n A'\\
            \forall i \leq k.\; \judge[\Sigma; \Gamma; f]{e_i}{\later^{m_i + j} A_i}
        }{
            \judge[\Sigma; \Gamma; f]{g(\overline{e})}{\later^{n + j} A'}
        }\and
        \inferrule[T-Case]{
            \judge[\Sigma; \Gamma; f]{e}{\later^n A}\\
            \calC_A = \{C_1, \dots, C_k\}\\
            \forall i \leq k.\; \sig(C_i) = (A_{i1}, \dots, A_{ik_i}) \\
            n \leq m\\
            \forall i \leq k.\; \judge[\Sigma; \Gamma, x_{i1} : \later^{n+1} A_{i1}, \dots, x_{ik_i} : \later^{n+1} A_{ik_i}; f]{e_i}{\later^m A'}
        }{
            \judge[\Sigma; \Gamma; f]{\case{e}{C_i}{x_i}{e_i}^k}{\later^m A'}
        }
    \end{mathpar}
    \caption{%
        Selected typing rules for the \tali language.
        A complete listing of the typing rules can be found in \autoref{apx:typeRules}.
    }
    \label{fig:typeRules}
\end{figure}
\subsection{Probabilistic inference for \tali Programs}
\label{subsec:probInference}

Since OBC \tali programs are guarded and productive by construction, every well-typed program reduces to weak-head normal form (WHNF) in finite time.
In our denotational semantics, this means that every well-typed program assigns zero probability to $\bot$, even if it does not terminate completely.
As a result, the likelihood of a given value can be computed as the sum of likelihoods over a finite set of demands.
Operationally, we implement probabilistic inference for \tali programs using lazy enumeration \citep{kiselyovEmbeddedProbabilisticProgramming2009,kollerEffectiveBayesianInference1997}. We construct thunks for each probabilistic draw and recursive function call, and force only when needed, such as in control flow.
Inference proceeds ``outside-in'' on a given value: we force the root constructor before its children and continue recursively in a top-down fashion.
Thunking probabilistic draws also allows for certain language extensions to be implemented easily, such as offsetting and clamping integer draws.

The inference procedure does not simply compute a numeric solution to the semantic equations in \autoref{fig:semantics}.
Instead, each forcing step produces one or more nodes in an arithmetic circuit that represents the semantic equations symbolically in terms of the probabilistic parameters.
This allows for efficient parameter tuning, which is exploited during synthesis (see \autoref{subsec:paramFitting}).
The OBC guarantees are sufficient for this to be feasible.
The arithmetic circuit for the probability mass that a program assigns to an observation can be decomposed into a sequence of recursive-call subcircuits.
Since the number of recursive calls is bounded by the size of the observation, the number of sub-circuits is similarly bounded, and since the set of demands at each recursive level is finite, each sub-circuit also has finite size.

\section{Learning Stochastic Summaries}
\label{sec:synthesis}

To automatically construct generative models of data, we use program synthesis to search for program structures with high marginal likelihood.
After identifying a likely model, we optimize its parameters to fit the data using maximum likelihood estimation (MLE).

\subsection{Synthesis of probabilistic programs}

To synthesize a probabilistic program that best describes a given dataset, we use a sketch-based stochastic synthesis algorithm.
Stochastic search and Monte Carlo techniques are well established tools for probabilistic program synthesis \citep{noriEfficientSynthesisProbabilistic2015,saadBayesianSynthesisProbabilistic2019}.
The inputs to the synthesis algorithm are a user-provided program sketch $\calP$ with one or more holes, and a training set of observations.
Given these inputs, the synthesis algorithm searches for a complete program $\calP[\overline{e}|\theta]$, with filled-in expressions $\overline{e}$ and parameter vector $\theta$ that maximize the likelihood of the observations.

\subsubsection{Likelihood evaluation and program scoring}
\label{subsec:likelihoodEval}

Program synthesis is driven by a likelihood-directed program scoring function.
Programs that explain the data well have high marginal likelihood, defined as the integral of the program's observation-conditional likelihood over the parameter space.
The \tali evaluator computes an arithmetic circuit for the likelihood of a set of observations under a given parameter assignment.
This makes marginal-likelihood computation fairly straightforward, though not always tractable.
In the special case where the likelihood circuit is a monomial, every observation is unambiguous---that is, a unique set of random choices during execution could have produced it.
For such circuits, the exact marginal likelihood can be computed in closed form with beta priors over each Bernoulli sample in the trace.
For general polynomial circuits, the marginal likelihood may be intractable, so outside of the monomial case we use mean-field variational Bayes to compute an evidence lower bound (ELBO) for the candidate program \citep{bleiVariationalInferenceReview2017,bealVariationalBayesianEM2003}.
One consequence of this approach is that the marginal-likelihood score favors unambiguous programs with monomial likelihoods.

Each program's score combines its marginal-likelihood score with the syntax-tree size of its hole fillings.
We also add a termination penalty for programs that are not almost-surely terminating.
Well-typed \tali programs are productive by construction, but this does not imply almost-sure termination; for example, a candidate programs may implement a supercritical branching process which diverges with nonzero probability.
Nonterminating programs are undesirable as generative models.
However, because our lazy-enumeration inference technique prunes paths based on the observations, it cannot distinguish a program that diverges on part of its support from one that terminates but fits the data poorly on that same part.
Thus we penalize each proposed candidate proportional to its divergence probability to down-weight non-terminating programs.
The divergence probability of an arbitrary program is uncomputable in general, so we approximate it empirically by drawing samples from the program using a fixed execution budget, and calculate the proportion of executions that exhaust the budget.

The scoring function $S$ for a given program $P$ with hole fillings $\overline{e}$ and parameter assignment $\theta$ is then
\[
  S(P[\overline{e}|\theta])=-\log\widehat{p}(\calD \mid P)
           +\log|\overline{e}|+T(P),
\]
where $|\overline{e}|$ is the total syntax-tree size of the hole fillings, $T$ is the empirical termination penalty, and $\log\widehat{p}$ denotes exact log evidence in the monomial case and its variational lower bound otherwise.

\subsubsection{Initial program search}
\label{subsec:initialProgSearch}

The synthesis procedure consists of an enumerative search over a feasible set of recursive program structures, followed by a Metropolis-style stochastic search guided by the likelihood of the observations under each model.
In our evaluations, we synthesize programs from sketches which contain at least one recursive function, with each entire function body left as a hole to be filled.

Initialization searches a restricted family of recursive programs for a promising starting program with nonzero likelihood on the training observations.
This is essential since the joint likelihood gives no partial credit for explaining only part of the dataset: omitting even one observed constructor assigns the dataset probability zero.
Most randomly generated bodies therefore receive the same infinite score, leaving pure stochastic search with no useful signal from the scoring function.
We instead search for an initial \emph{covering skeleton}: an initial hole filling which assigns non-zero probability to every observation.

For a given output type, a basic skeleton contains a cascade of probabilistic choices with one alternative per observed constructor.
Inductive constructor fields are filled by recursive calls, integer and boolean fields by primitive samples, and flat fields by a choice over their enumerated constructors.
For instance, consider a recursive Boolean list generator over constructors $\con{Nil}$ and $\con{Cons}$.
Its basic skeleton consists of a single-choice cascade covering $\con{Nil}$ and $\con{Cons}$, where the $\con{Cons}$ fields are filled with a primitive $\flip$ and a recursive call for its tail.

A sketch typically has one generator per type, so every call to it draws from the same constructor distribution, regardless of where in the tree the call occurs.
This cannot express many context-sensitive structures, such as a tree where children can be drawn from different distributions depending on the parent's label.
Before building the basic skeletons, we optionally split each generator by call-site, following parent annotation \citep{johnsonPCFGModelsLinguistic1998}.
A call site is either the root of an observation or a constructor field (for example, the left child of a $\con{Node}$).
Because each subterm's call site is visible in the data, we can quickly compute the skeleton's marginal likelihood in closed form from per-site constructor counts and field values under Beta priors, without running the full evaluator.
We begin with one class per site, and, following Bayesian model merging \citep{stolckeInducingProbabilisticGrammars1994,hwangInducingProbabilisticPrograms2011}, greedily merge classes by marginal likelihood, stopping when no candidate merger would increase it.
Each remaining class becomes a constructor of a new tag type, which is passed to the skeleton as an additional argument.
The function takes this tag as an additional argument and matches on it, producing an independent sub-skeleton cascade for each tag constructor.
Each cascade arm passes the tag as an argument to its recursive calls.

Within each skeleton, we choose an expression for each integer field and the arguments for each recursive call.
The grammar includes multiple forms for integer expressions, such as geometric draws (optionally offset or clamped) and mixtures over point masses and geometric draws; the full skeleton grammar is given in \autoref{apx:synthDSL}.
Generated fields are bound in $\kw{let}$ expressions so that their values may be shared in subsequent expressions.
Each field therefore has a finite set of choices that can be enumerated during search.

These choices must often be made jointly.
To illustrate, consider a sketch for a sorted-list generator $\fn{sorted}(l: \mathsf{Int}, h: \mathsf{Int})$, where the arguments are the current lower and upper bounds.
Generating the head $x = \fn{min}(\geom \theta + l, h)$ and passing $x$ as the lower bound in the recursive call $\fn{sorted}(x,h)$ produces an ordered list within the interval.
These choices must be made together, and the variable $x$ must be threaded correctly.
Searching these choices jointly exposes useful dependencies that isolated subtree changes might not discover.

The product of this choice space is finite but can be large.
We fix an evaluation budget; if it exceeds the size of the choice space, we enumerate the space exhaustively.
Otherwise, we enumerate field forms and perform coordinate descent over call arguments for each combination.
If this is still too expensive, we perform coordinate descent over all choices, starting from the default seed with bare draws and default argument passing.
Each sweep retains strictly improving changes and stops when a fixed point is reached.
The best-scoring seed initializes the chain, and we retain the evaluated assignments for restarts (see below).
This search offers no guarantee of global optimality when enumeration is incomplete.

\subsubsection{Stochastic search}
\label{subsec:stochasticsSearch}

Starting from the selected seed, stochastic search explores the program space using a subtree replacement kernel.
Each generated subtree records its target type, in-scope variables, guardedness, and outstanding relevance obligations.
A replacement is generated in that saved context with fresh local binders, preserving the expression's type and variable bindings.
Per the type system, recursive self-calls are permitted only in constructor-guarded positions.
Subtrees must also satisfy a variable-use requirement, following relevant typing \citep{pierceAdvancedTopicsTypes2005}: each inductive function argument must be used at least once in every control-flow branch.
The generator chooses among admissible typing rules and fills their subexpressions recursively, backtracking when a rule's obligations cannot be satisfied.
Consequently, the search explores a restricted subset of well-typed \tali programs which expresses a variety of recursive structures.

The stochastic search mixes structural and local proposals.
At each step, the search selects either structural sites (conditionals, matches, constructor applications, and $\kw{let}$ bindings) or argument sites (calls, primitive operations, and atomic expressions).
Sites are selected uniformly within the chosen class; if that class is empty, the search may use any non-root replacement site.
If local replacement fails, the selected hole is regenerated wholesale.
The probability of choosing a structural site rather than an argument site decays during the search.
Thus, early proposals more often revise branching and sharing structure, while later proposals emphasize data flow without eliminating structural exploration.

To avoid getting stuck in local minima, the search can periodically regenerate a hole's entire filling.
Fresh restarts are drawn without replacement from the searched seed assignments, weighted by their scores.
Alternatively, the search can make a fresh type-directed draw for the hole.

Each proposed program candidate is scored with the scoring function $S$ described in \autoref{subsec:likelihoodEval}.
For finite scores, a proposed program $P'$ replaces the current accepted program $P$ with probability
\[
  A(P,P')=
  \min\left\{1,\exp\left(S(P)-S(P')\right)\right\}.
\]
Proposals with infinite score are rejected from a finite-score state; when both scores are infinite, the proposal is accepted to permit continued exploration.
Because the proposals are asymmetric and change with the search schedule, and because the acceptance rule omits a proposal-ratio correction, this is \emph{not} a posterior sampler over the program space.
Rather, it is a Metropolis-style stochastic search for program selection.

\subsubsection{Parameter fitting}
\label{subsec:paramFitting}

After stochastic search is complete, we fit parameters for the top-$k$ program structures found during the search.
Since the inference algorithm constructs an arithmetic circuit representing the likelihood of the observations given the program parameters, we are able to compute gradients of the likelihood function and optimize the parameters using L-BFGS, a gradient-based optimization algorithm \citep{liulimited1989}.
After optimization, each of the top-$k$ programs is scored against the validation set, and the best program is returned.

\section{Evaluation}
\label{sec:evaluation}

We evaluate \tali on a synthetic benchmark grid with data generated from known ground truth programs and on two held-out real-world domains, regular expressions \citep{davisWhyArentRegular2019} and SPARQL queries \citep{malyshevGettingMostOut2018}.
We hypothesize that, with tens to hundreds of training trees, the structural prior of the OBC program class can allow synthesized programs recover the structure of a distribution from far fewer examples than data-intensive models such as transformers.

We examine the following research questions in our evaluation:

\noindent
\textbf{RQ1:} How well do synthesized programs fit the data distribution, compared with matched-data neural baselines, and how does this change with training-set size in the low-data regime?\\
\textbf{RQ2:} Which invariants of the data do synthesized programs recover, and how much training data is needed for reliable learning?\\
\textbf{RQ3:} Can synthesized programs act as a data augmentation source to improve the performance of downstream model training tasks on real-world data?\\

\noindent

\subsection{Benchmarks}
\label{subsec:benchmarks}

\begin{table}[t]
  \centering
  \scriptsize
  \renewcommand{\arraystretch}{1.25}
  \begin{tabular}{@{}p{0.105\textwidth}p{0.25\textwidth}
                      p{0.25\textwidth}p{0.25\textwidth}@{}}
  \hline
  & \textbf{String (D1)} & \textbf{Tree (D2)} & \textbf{Expr.\ AST (D3)} \\
  \hline
  \textbf{Geometric (S1)}
  & Stop, or emit a symbol and recurse.
  & Labelled leaf, or branching node.
  & Literal, or recursive operator. \\
  \textbf{Fuel (S2)}
  & Fuel bounds length; early stop allowed.
  & Fuel bounds depth; children get less.
  & Fuel bounds depth across both sorts. \\
  \textbf{State (S3)}
  & Latent state drives stopping and symbols.
  & Latent state drives stopping and leaves; transitions label children.
  & Latent state drives literals and operators; transitions follow child position. \\
  \textbf{Constraint (S4)}
  & Inherited interval; each symbol raises the floor.
  & Inherited interval; a pivot splits it between children.
  & Inherited scope removes operators; some tighten it. \\
  \hline
  \textbf{Local (V1)}
  & Each symbol within $c$ of the previous.
  & Each key within $c$ of its parent's.
  & Literals within $c$ of the enclosing operator's value. \\
  \textbf{Interval (V2)}
  & Sorted list (as S4).
  & Search tree (as S4).
  & Variable indices name an enclosing binder. \\
  \textbf{Aggregate (V3)}
  & Head holds the list's length.
  & Each node holds its subtree's size.
  & Each operator holds the sum of its operands. \\
  \hline
  \end{tabular}
  \caption{Synthetic benchmark grid. Rows vary recursive structure (S) and value relation (V), columns vary data domain.}
  \label{tbl:syntheticGrid}
\end{table}

\subsubsection{Synthetic grid}
The grid (\autoref{tbl:syntheticGrid}) comprises 21 ground-truth program grammars over three data domains: strings, binary trees, and expression ASTs.
The \emph{structure} axis (S1--S4) varies how the ground truth recurses; the \emph{value-relation} axis (V1--V3) varies how far a dependency between values reaches: to a neighbor, to every ancestor, or to an aggregate of a whole subtree.
S1--S4 and V2 are naturally expressible as OBC programs; V1 requires threading a value through the recursion, and V3 depends on subtrees not yet generated.
Known ground truths let us measure fit as excess NLL over the truth, and validity as membership in the truth's support.

\subsubsection{Real-world data}
The regex dataset is derived from a corpus collected in the empirical software engineering literature \citep{davisWhyArentRegular2019}; the SPARQL dataset is derived from a collection of queries to Wikidata~\citep{malyshevGettingMostOut2018}.
We deduplicated, normalized and converted both to ASTs, yielding over 414,000 and 148,000 unique syntax trees, respectively.
Each regex character is encoded by its class and its index within that class, so a program must learn the exact distribution over the characters of fixed strings.
Regex results use uniform samples of the corpus, except for rules about rare constructs (bounded repeats, set ranges, backreferences), which we measure on training sets drawn from regexes containing those constructs.

\subsection{Experimental Setup and Baselines}
\label{subsec:baseline-description}

Our method fits a program to each set of $n$ training trees using the synthesis algorithm of \autoref{sec:synthesis}.
We then pretrain a transformer on 10,000 samples from the synthesized program and fine-tuned on the original $n$ trees (\emph{program-pretrained}).
We compare against two transformers trained on the same $n$ trees: one from scratch (\emph{no pretraining}), and one first pretrained on 10,000 random well-typed values (\emph{random-pretrained}). 
As references, we use the ground truth on the grid and, on real data, a transformer trained on the entire training pool (\emph{full pool}), which sees orders of magnitude more data than any other approach.
We use $n \in \{10, 30, 100, 300\}$ and 30 seeds per setting (27 for regex), and report medians across seeds, means for rates (validity, violation and recall), and \emph{win fractions}: the fraction of seeds on which one arm beats another on the same training set.

\subsubsection{Fit metrics}
A synthesized program can assign a test tree probability zero, so for a program $p$ and comparison model $r$ we report the paired excess NLL over the test trees $X_p$ that $p$ covers:
\[
  \frac{1}{|X_p|}\sum_{x \in X_p} \bigl(-\log \dbrack{p}(x) + \log r(x)\bigr), \qquad X_p = \{x \in X : \dbrack{p}(x) > 0\}
\]
Alone, this can overstate the fit of a program that covers few trees; as an extreme example, consider a program which assigns very high likelihood to exactly one test tree, but zero likelihood to the rest.
Such a program would have perfect \emph{covered} excess NLL, despite severely overfitting.
To mitigate this, we always report excess NLL beside \emph{support recall} and, for structure, \emph{support precision}, where $q$ is the data distribution:
\[
  \text{recall}(p, q) = \Pr_{x \sim q}\left(p(x) > 0\right)%
    \qquad \text{prec}(p, q) = \Pr_{x \sim p} \left(q(x) > 0\right)
\]
We estimate recall as the fraction of test trees with finite NLL; transformers have recall 1 by construction.

\subsubsection{Structure metrics}
Invariants are a matter of precision: a model that ``knows'' lists are sorted never emits an unsorted list.
On the grid, we score 1,000 samples per model under the ground truth.
Every model, the truth included, often replays training trees (55--79\% of the truth's samples), which are valid by construction, so we report violation rates on \emph{novel} samples only, and count transformers' malformed outputs (token sequences that do not decode to a tree) separately.
For real data, we check domain validity rules instead: for regexes, compilation by Python's \texttt{re} module, $m \leq n$ in bounded repeats \texttt{\{m,n\}}, ordered set-range bounds, and backreferences to closed groups; for SPARQL, well-formedness and four scoping rules from the specification (\autoref{fig:structureRules}).

\subsubsection{Held-out data}
The grid was our development set.
The real-world experiments were preregistered after the method was frozen; we added the program-pretraining arm and the regex matched-$n$ baselines before examining any results.
Three regex seeds informed synthesizer design and are excluded from all reported results.
SPARQL was never examined during development.

\smallskip
We evaluate RQ1 and RQ2 separately for the the synthetic dataset and real-world datasets in Sections~\ref{subsec:learnSynthGrid} and~\ref{subsec:learnRealWorld}. We then dedicate Section~\ref{subsec:dataAugmentation} to addressing RQ3.

\subsection{Learning Programs over the Synthetic Grid}
\label{subsec:learnSynthGrid}

\begin{figure}[t]
    \centering
    \includegraphics[width=0.7\linewidth]{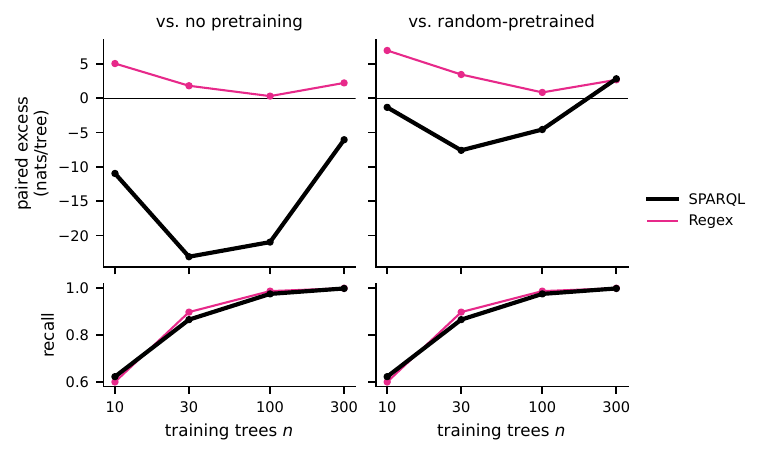}
    \caption{%
      Median paired excess NLL of synthesized programs versus matched-$n$ transformers, and mean program support recall on the test set (across seeds).
      Left: Paired excess NLL and recall of the synthesized program versus the no-pretraining model.
      Right: Paired excess NLL and recall of the synthesized program versus the random-pretraining model.
    }
    \label{fig:fitRecall}
\end{figure}

The results of the evaluation on the synthetic grid are in \autoref{tbl:grid}.
Each cell in the grid corresponds to a set of ground truth programs, some of which are easily expressed in the synthesis DSL and others which are less so.
In cases where the synthesis DSL can directly express the ground truth program, programs are more sample-efficient models than either transformer baseline at the same training set sizes .
As a particularly stark example, the program's median excess over the truth on cells S1--S3 falls to nearly zero at only $n=100$: it effectively recovers the exact generating distributions, which are variants on straightforward and productive ``sample-then-recurse'' patterns.
On many grid cells with additional constraints, such as S4 and V2 which enforce value orderings, its excess is only a third to two fifths of the random-pretrained transformer's at $n=300$, and it fits better on 74--93\% of runs at $n \geq 100$.

Probabilistic programs also typically achieve high recall across all grid cells.
At $n=300$, programs reach median recall greater than 99\% on 10 of the 21 cells.
This is partly an effect of the skeleton seeding process; a skeleton has full recall by construction, and stochastic search generally stays within a high-recall region of candidate programs.

In the very-low data regime, program synthesis is able to cover a large portion of the support, though its fit is not much better than the baselines.
At $n=10$, the program misses 7--12\% of test trees and is only level with random pretraining on the rest.

Constraint recovery on the synthetic grid, much like model fit, is highly correlated with proximity to the OBC class and the synthesis DSL.
The OBC property requires values to be generated top-down: local constraints are easy to express, while long-distance dependencies across different parts of a value are not.
The synthesis DSL and skeleton seeding (\autoref{subsec:initialProgSearch}) reflect this bias, and parent annotation adds local Markov-style context to each choice.
The results follow this boundary closely: where the class can express an invariant, programs learn it from relatively few examples and maintain it better than matched-$n$ transformers.
The \emph{structure} columns of \autoref{tbl:grid} (novel violation, precision) give the result.
The clearest example is the V2 row, containing sorted lists, search trees, and scoped variables.
When learning interval invariants, for instance, the transformer must infer these constraints from data alone, while the program can encode them directly. 
Programs violate constraints on 13--20\% of novel samples, substantially lower than the untrained and random-pretrained transformers.
On the other hand, the synthesis generally fails to learn constraints which are difficult for OBC programs or the synthesis DSL to express.
On V1 and V3, for instance, the program loses to both transformers at $n=30$ and higher: V3 cannot easily be expressed top-down, and the V1 program that threads a value through the recursion is rarely found.

\subsection{Learning Programs over Real-World Data}
\label{subsec:learnRealWorld}

\begin{table}[t]
  \centering
  \scriptsize
  \setlength{\tabcolsep}{4pt}
  \begin{tabular*}{\linewidth}{@{\extracolsep{\fill}}lrrrrrrrrrrrr}
\toprule
 &  & \multicolumn{3}{c}{excess over truth} & \multicolumn{2}{c}{fit wins} &  & \multicolumn{3}{c}{novel violation} & \multicolumn{2}{c}{precision wins} \\
\cmidrule(lr){3-5}\cmidrule(lr){6-7}\cmidrule(lr){9-11}\cmidrule(lr){12-13}
Block & $n$ & P & R & N & vs.\ R & vs.\ N & recall & P & R & N & vs.\ R & vs.\ N \\
\midrule
S1--S3 & 10 & $\mathbf{0.35}$ & $0.38$ & $2.58$ & .43 & .67 & .893 & $0.03$ & $\mathbf{0.02}$ & $0.33$ & .34 & .92 \\
 & 30 & $\mathbf{0.16}$ & $0.29$ & $1.23$ & .56 & .77 & .970 & $0.02$ & $\mathbf{0.02}$ & $0.19$ & .36 & .88 \\
 & 100 & $\mathbf{0.03}$ & $0.21$ & $0.56$ & .82 & .93 & .995 & $\mathbf{0.01}$ & $0.02$ & $0.09$ & .46 & .79 \\
 & 300 & $\mathbf{0.01}$ & $0.12$ & $0.25$ & .91 & .97 & 1 & $\mathbf{0.01}$ & $0.02$ & $0.06$ & .47 & .74 \\
\addlinespace
S4 & 10 & $\mathbf{0.93}$ & $1.12$ & $3.89$ & .43 & .62 & .904 & $\mathbf{0.20}$ & $0.30$ & $0.43$ & .66 & .75 \\
 & 30 & $\mathbf{0.55}$ & $0.92$ & $2.01$ & .62 & .71 & .975 & $\mathbf{0.22}$ & $0.33$ & $0.34$ & .72 & .66 \\
 & 100 & $\mathbf{0.20}$ & $0.71$ & $1.10$ & .79 & .77 & .997 & $\mathbf{0.22}$ & $0.36$ & $0.27$ & .74 & .65 \\
 & 300 & $\mathbf{0.13}$ & $0.40$ & $0.58$ & .74 & .72 & 1 & $0.28$ & $0.33$ & $\mathbf{0.22}$ & .59 & .48 \\
\addlinespace
V2 & 10 & $\mathbf{0.41}$ & $1.05$ & $3.77$ & .59 & .87 & .909 & $\mathbf{0.15}$ & $0.35$ & $0.47$ & .83 & .86 \\
 & 30 & $\mathbf{0.44}$ & $0.89$ & $2.22$ & .73 & .92 & .965 & $\mathbf{0.14}$ & $0.35$ & $0.39$ & .87 & .84 \\
 & 100 & $\mathbf{0.29}$ & $0.81$ & $1.31$ & .89 & .93 & .992 & $\mathbf{0.13}$ & $0.36$ & $0.34$ & .89 & .84 \\
 & 300 & $\mathbf{0.21}$ & $0.51$ & $0.75$ & .93 & .90 & .998 & $\mathbf{0.20}$ & $0.42$ & $0.34$ & .81 & .72 \\
\midrule
V1 & 10 & $\mathbf{0.49}$ & $1.37$ & $3.45$ & .56 & .74 & .877 & $\mathbf{0.54}$ & $0.60$ & $0.62$ & .49 & .54 \\
 & 30 & $1.15$ & $\mathbf{0.93}$ & $2.17$ & .38 & .61 & .958 & $0.67$ & $0.57$ & $\mathbf{0.49}$ & .35 & .28 \\
 & 100 & $1.26$ & $\mathbf{0.67}$ & $1.08$ & .14 & .34 & .994 & $0.80$ & $0.39$ & $\mathbf{0.35}$ & .07 & .03 \\
 & 300 & $1.22$ & $\mathbf{0.46}$ & $0.64$ & .03 & .11 & .999 & $0.87$ & $0.32$ & $\mathbf{0.31}$ & .01 & .01 \\
\addlinespace
V3 & 10 & $\mathbf{2.32}$ & $3.08$ & $5.37$ & .63 & .88 & .933 & $\mathbf{0.71}$ & $0.72$ & $0.72$ & .57 & .40 \\
 & 30 & $\mathbf{2.10}$ & $2.74$ & $3.14$ & .65 & .72 & .982 & $0.76$ & $\mathbf{0.75}$ & $0.75$ & .44 & .34 \\
 & 100 & $1.99$ & $2.16$ & $\mathbf{1.77}$ & .57 & .36 & .995 & $0.83$ & $0.81$ & $\mathbf{0.64}$ & .49 & .23 \\
 & 300 & $1.93$ & $1.26$ & $\mathbf{0.64}$ & .39 & .32 & .999 & $0.89$ & $0.77$ & $\mathbf{0.56}$ & .26 & .19 \\
\bottomrule
\end{tabular*}

  \caption{%
    The synthetic grid, by block and training-set size $n$, for the synthesized program (P), the random-pretrained transformer (R), and the untrained transformer (N).
    \emph{Fit:} median excess NLL over the ground truth (nats per tree), over the test trees the program covers for P and over all test trees for R and N; the fraction of runs on which the program fits better than each transformer, paired on the trees it covers; and the program's mean support recall.
    \emph{Structure:} mean rate at which novel, well-formed samples fall outside the ground truth's support; and the fraction of runs on which the program's novel samples do so less often than each transformer's (ties count as losses).
    The best excess and violation rate in each row are in bold; win fractions above .5 favor the program.
  }
  \label{tbl:grid}
\end{table}

Synthesizing probabilistic programs which fit real-world data is a more challenging learning task. Of course, we cannot expect programs to fit these distributions perfectly. Instead, we evaluate the recovery of structural constraints present in the dataset, focusing on syntactic and semantic validity of the trees representing the regexes and SPARQL queries. 
We find that synthesized programs achieve reasonable learning performance in certain respects, though they are not especially competitive as likelihood estimators.
On the regex dataset, for instance, the program is roughly level on fit with the transformers at $n=100$, though it is generally behind at $n=300$.
(It should be noted that program fit on regexes is dominated by a lexical task, predicting the individual characters of fixed strings.
When characters within a class are treated uniformly, the synthesized program actually beats the untrained transformer at $n=100$ on 67\% of seeds, and on the bounded-repeat subset, where structure recovery is paramount, it beats both transformers at every $n \geq 30$, sometimes uniformly across seeds.)

For regexes, the program's value lies chiefly in the structure it supplies as a teacher (see \autoref{subsec:dataAugmentation}) and in the validity of its samples, more than as an accurate likelihood estimator.
Programs produce compiling regexes in 51--54\% of samples at every $n$, against 0.3--5\% for the untrained transformer and 10--24\% for the random-pretrained one.
The pattern of good local constraint recovery we note on the synthetic grid continues on the regex set: local rules, such as bounded repeat  or character range orderings, are almost always respected.
At $n=300$, for instance, only 0.5\% of  bounded repeats \texttt{\{m,n\}} in program samples have $m > n$, against 9.3\% and 30.7\% for the transformers.
Furthermore, only 0.5\% of character ranges are reversed (e.g. \texttt{[z-a]}).
On the other hand, long-range dependencies are typically not recovered; 77\% of backreferences in program samples name an unclosed group compared to 25--39\% for the transformers.

Programs fit to the SPARQL dataset also function better as structural teachers than as accurate likelihood estimators, though in that role it functions better than the baselines.
Indeed, the program is the best density model at $n=30$ and $n=100$ (\autoref{fig:fitRecall}).
At $n=300$ it still beats the untrained transformer on every seed, but its relative advantage diminishes, as random pretraining overtakes it slightly.

As with the regex dataset, the structural recovery by synthesized programs is much stronger than the baselines on the SPARQL dataset.
Across all seeds, 40--48\% of program samples are valid queries, against 0--2\% for no-pretraining and 3--21\% for random-pretraining.
The program is more valid than both transformers on every seed at every $n$, and already at full strength at $n=10$.
Similarly to regex, long-range SPARQL dependencies are often not recovered.
Synthesized programs break SPARQL grouping rules, \texttt{BIND} scope restrictions, \texttt{AS} scope and variable ordering constraints in up to 99\% of applicable samples.
The full-pool transformer learns all of these rules (at worst 12\% violations), indicating the failure reflects the program class rather than the data.

Parent annotation provides most of the benefit that programs enjoy on real data.
Without it, only 2--3\% of program samples are valid SPARQL queries, because the root query pattern is drawn from the same rule as every nested one.
Annotation also has a cost on the grid: it gives up the exact V2 recovery of the unannotated program in exchange for better fit.

\begin{figure}[t]
    \centering
    \includegraphics{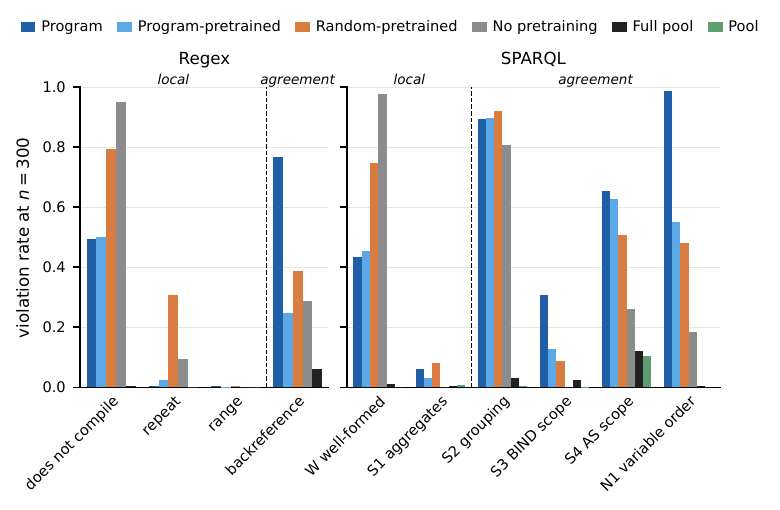}
    \caption{%
      Fraction of samples violating each validity rule at $n=300$ (mean over seeds; lower is better), grouped into \emph{local} rules (placement and order of one construct) and \emph{agreement} rules (relating two uses of a variable or group).
      \emph{Regex:} fails \texttt{re.compile}; repeat \texttt{\{m,n\}} with $m > n$; set range with reversed bounds; backreference to an unclosed group.
      \emph{SPARQL:} not printable as a query (W); misplaced or nested aggregate (S1); aggregate query projects a non-\texttt{GROUP BY} variable (S2); \texttt{BIND} to an in-scope variable (S3); \texttt{AS ?v} rebinds a bound variable (S4); variables not numbered in first-use order (N1).
      Except for compilation and W, rates are among samples containing the rule's construct.
      Pool: the SPARQL training data, which includes some queries the specification forbids.
    }
    \label{fig:structureRules}
\end{figure}

\subsection{Data Augmentation with Synthesized Programs}
\label{subsec:dataAugmentation}

\begin{figure}[t]
  \includegraphics{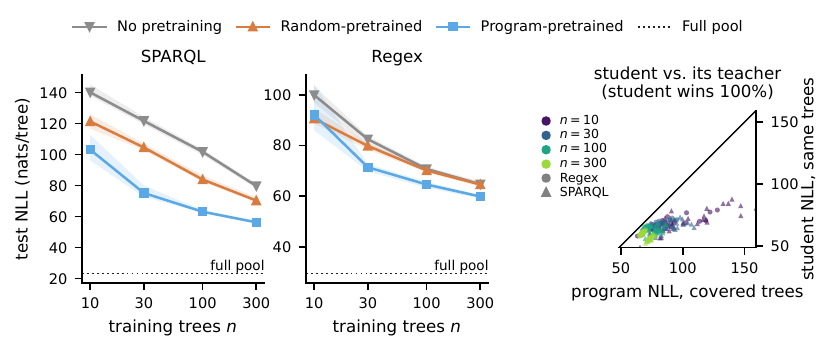}
  \caption{%
    Left: Test NLL for transformers trained without pretraining, pretrained on random well-typed values, or pretrained on samples from a synthesized program, with 10--300 training trees (median and interquartile range across seeds); the dotted line shows a transformer trained on the full pool.
    Right: For each run on both domains, the synthesized program's NLL on the test trees it covers versus the NLL of the transformer pretrained on samples from that program.
    Every point lies below the diagonal: on the covered trees, the student has lower NLL than its teacher.
  }
  \label{fig:useLearnCurve}
\end{figure}

In this section we evaluate whether the data-efficient summaries we learn can serve as effective teachers for downstream learning tasks. We do this by comparing the performance of the program-pretrained transformers against other transformers that were not fine-tuned on synthetically generated data from our learned programs.%
A synthesized program is a sampler, so it can amplify what it learns from only $n$ trees into an arbitrarily large corpus.
We find that pretraining on the program-sampled corpus is in fact the best of the three transformer pretraining strategies on both real-world domains at every $n \geq 30$ (\autoref{fig:useLearnCurve}, left).
The program-pretrained transformer beats the random-pretrained one on every seed at $n \geq 30$ on SPARQL and at $n \geq 100$ on regex.
The advantage comes from the program itself, rather than from pretraining in general: for instance, only 12\% of the samples from the random-pretrained regex model actually compile, against 46\% of program-pretrained samples.

Notably, the program-pretrained transformer fits the trees the program covers better than the program itself on every run in both real-world domains (\autoref{fig:useLearnCurve}, right).
At $n=300$, the program-pretrained model is a better-fit model of the SPARQL dataset on all 30 seeds, and by a substantial margin (a median of 17.5 nats).
The program reinforces the structural information which is encoded the student model by introducing novel samples which reify the constraints present in the training set, and fine-tuning locally corrects some of the program's residual.
The full-pool transformer, trained on over 100,000 trees, remains a much better model, but at only 300 training trees, program pretraining closes 41\% of the gap between the untrained transformer and the reference on SPARQL, and 14\% on regex.

\subsection{Discussion and Limitations}

Within the OBC class, synthesized programs recover the structure of a distribution from tens to hundreds of examples: they keep its invariants more reliably than transformers trained on the same data (RQ2), fit it more closely on the grid and on SPARQL up to $n=100$, (RQ1), and transfer that structure to a neural model through pretraining (RQ3).
Outside the class, on agreement rules and aggregate values, they fail predictably, and the full-pool transformer shows that these failures reflect the class rather than the data.
Below $n=30$, programs omit constructs absent from the training set, so they are useful there as pretraining sources and for their validity, but not as stand-alone density models.
We believe that with better program synthesis techniques, a more expressive synthesis DSL, and perhaps more advanced methods such as ensemble learning or specification mining, some of the gaps on real-world data could be closed, especially in the data augmentation setting.

No transformer uses constrained decoding \citep{gengGrammarConstrained2023}, which biases their real-data violation rates upward, since malformed outputs count as invalid; on the grid, we report precision on well-formed samples only, so those results are unaffected.
We believe the choice is justified: with enough data, malformed outputs nearly disappear (99.6\% of the full-pool transformer's regex samples compile, and 96.6\% of its SPARQL samples are valid), and constrained decoding would make the likelihoods a model assigns correspond to a different distribution from the one it samples.

\section{Related work}

\paragraph{Operational Profiling}

Operational profiling \citep{musaOperationalProfilesSoftwarereliability1993,gittensExtendedOperationalProfile2004} is an approach to software evaluation which emphasizes testing and measuring software systems within the context they are designed to be used or are currently deployed.
Operational profiling involves both quantitative and statistical methods to characterize typical usage \cite{hartmannStatisticalAnalysisOperational2016,dubousquetExpressingImplementingOperational1998}, as well as qualitative and user-centric methods to develop user stories and contexts to improve software reliability \citep{zekiciTestingSoftwareOperational2000,aroraSoftwareReliabilityImprovement2005,smidtsSoftwareTestingOperational2014}.
Operational profiling seeks to understand empirically how software behaves when situated in its operational context, and subsequently test software in that context.
In this vein, \citet{paveseInputs2018} study the problem of learning PCFGs over real-world program inputs for use in robustness testing.
Our technique is similar to this approach but uses a more expressive hypothesis space.

\paragraph{Probabilistic Program Synthesis}

Prior work has explored techniques to synthesize probabilistic programs which fit datasets or satisfy specifications.
Some existing approaches use MCMC methods to synthesize programs in various DSLs which maximize the likelihood of an observed dataset \citep{saadBayesianSynthesisProbabilistic2019,noriEfficientSynthesisProbabilistic2015}.
\textsc{DreamCoder} \citep{ellisDreamCoderBootstrappingInductive2021} is another synthesis-driven learning tool which learns representations using programs.
Learning programs as representations is a standard inductive bias in machine learning~\citep{muggletonInductiveLogicProgramming1994}.
We frame our tool as as a framework which learns a distribution over tree-structured values using probabilistic program synthesis; as a representation-learning approach, we produce a program as our representation.

\paragraph{Probabilistic Programming Languages}

Probabilistic programming languages have long been a foundational tool in generative modeling contexts~\citep{goodmanChurchLanguageGenerative2014,gordonProbabilisticProgramming2014,claretBayesianInferenceUsing2013,ghahramaniProbabilisticMachineLearning2015}.
While probabilistic programming is a powerful computational paradigm, inference is often inefficient or intractable.
Recent research has made great strides in efficient discrete probabilistic inference.
Languages such as \texttt{Dice} \citep{holtzenScalingExactInference2020} exploit knowledge compilation to transform discrete probabilistic programs into binary decision diagrams, allowing efficient inference on a wide variety of programs.
Loaded Dice \citep{tjoaTuningRandomGenerators2025} builds upon Dice by adding automatic differentiation on the computation graph induced by the BDD representation of a probabilistic program for use in parameter optimization.
We take a very similar gradient-optimization approach for our parameter learning mode, though Loaded Dice is more focused on optimizing synthetic objectives such as dataset entropy.

Previous work has explored exact inference on recursive probabilistic programs \citep{chiangExactRecursiveProbabilistic2023,bowersStochasticLazyKnowledge2025,liCompilingProbabilisticPrograms2024,stuhlmullerDynamic2012}.
Many probabilistic programming languages use lazy strategies to achieve fast marginalization and inference \citep{bowersStochasticLazyKnowledge2025,narayanan2016probabilistic,kiselyovEmbeddedProbabilisticProgramming2009,scibiorFunctional2018,Pfeffer2015LazyFI}.
Languages such as Pluck \citep{bowersStochasticLazyKnowledge2025} which support recursive programs over inductive datatypes exploit the fact that expressions can naturally be evaluated to \emph{weak-head normal form}, where the forced constructor can be thought of as an observation which prunes the value space to be enumerated.
Our guarded recursion scheme forces recursive calls to only occur where the result of a recursive call will be the child of a constructor in WHNF, meaning that recursive calls can always be thunked.

The exactness guarantee of OBC differentiates our work from existing lazy inference techniques, which are sound but not necessarily convergent.
Figaro's lazy factored inference \citep{Pfeffer2015LazyFI}, for example, returns upper and lower bounds on probabilities which are not guaranteed to converge to a single value.
Pluck \citep{bowersStochasticLazyKnowledge2025} dynamically unrolls recursive programs under its lazy inference procedure, but does not guarantee that exact inference bounds the number of probabilistic choices.
Indeed, \citet{bowersStochasticLazyKnowledge2025} note that ``[i]t would be interesting to explore whether there are sensible syntactic restrictions that make such programs inexpressible;'' we partially answer this challenge.

\paragraph{Modal Types and Guarded Recursion}

Modal types for recursion were first described by \citet{nakanomodality2000a}, have been used to prove productivity in corecursive programs \citep{atkeyProductive2013}.
Guarded recursion schemes have been applied to tree-like structures \citep{birkedalFirst2011}, and have been used to reason about probabilistic computation and non-determinism \citep{stassenModelling2025a}, as well as almost-sure termination \citep{gregersenAlmostSure2024}.
Our approach draws from this work in order to prove an eventual stabilization property (OBC) about unrolled probabilistic programs.

\bibliographystyle{ACM-Reference-Format}
\bibliography{problearning}

@inproceedings{apellanizImproved2024,
  title     = {An {{Improved Tabular Data Generator}} with {{VAE-GMM Integration}}},
  booktitle = {2024 32nd {{European Signal Processing Conference}} ({{EUSIPCO}})},
  author    = {Apellaniz, Patricia A. and Parras, Juan and Zazo, Santiago},
  year      = 2024,
  month     = aug,
  pages     = {1886--1890},
  issn      = {2076-1465},
  doi       = {10.23919/EUSIPCO63174.2024.10715230}
}

@inproceedings{aroraSoftwareReliabilityImprovement2005,
  title     = {Software Reliability Improvement through Operational Profile Testing},
  booktitle = {Annual {{Reliability}} and {{Maintainability Symposium}}, 2005. {{Proceedings}}.},
  author    = {Arora, S. and Misra, R.B.},
  year      = 2005,
  month     = jan,
  pages     = {621--627},
  issn      = {0149-144X},
  doi       = {10.1109/RAMS.2005.1408433}
}

@inproceedings{atkeyProductive2013,
  title     = {Productive Coprogramming with Guarded Recursion},
  booktitle = {Proceedings of the 18th {{ACM SIGPLAN}} International Conference on {{Functional}} Programming},
  author    = {Atkey, Robert and McBride, Conor},
  year      = 2013,
  month     = sep,
  pages     = {197--208},
  publisher = {ACM},
  address   = {Boston Massachusetts USA},
  doi       = {10.1145/2500365.2500597},
  isbn      = {978-1-4503-2326-0},
  langid    = {english}
}

@inproceedings{barbalho2023virtual,
  title     = {Virtual Machine Allocation with Lifetime Predictions},
  booktitle = {{{MLSys}}},
  author    = {Barbalho, Hugo and Kovaleski, Patricia and Li, Beibin and Marshall, Luke and Molinaro, Marco and Pan, Abhisek and Cortez, Eli and Leao, Matheus and Patwari, Harsh and Tang, Zuzu and Santos, Tamires and Gon{\c c}alves, Larissa Rozales and Dion, David and Moscibroda, Thomas and Menache, Ishai},
  year      = 2023,
  month     = jun
}

@incollection{bealVariationalBayesianEM2003,
  title      = {The {{Variational Bayesian EM Algorithm}} for {{Incomplete Data}}: {{With Application}} to {{Scoring Graphical Model Structures}}},
  shorttitle = {The {{Variational Bayesian EM Algorithm}} for {{Incomplete Data}}},
  booktitle  = {Bayesian {{Statistics}} 7},
  author     = {Beal, Matthew J and Ghahramani, Zoubin},
  editor     = {Lindley, V and Bernardo, J M and Bayarri, M J and Berger, J O and Dawid, A P and Heckerman, D and Smith, A Fm and West, M},
  year       = 2003,
  month      = jul,
  pages      = {453--463},
  publisher  = {Oxford University PressOxford},
  doi        = {10.1093/oso/9780198526155.003.0025},
  isbn       = {978-0-19-852615-5 978-1-383-02433-3},
  langid     = {english}
}

@inproceedings{birkedalFirst2011,
  title      = {First {{Steps}} in {{Synthetic Guarded Domain Theory}}: {{Step-Indexing}} in the {{Topos}} of {{Trees}}},
  shorttitle = {First {{Steps}} in {{Synthetic Guarded Domain Theory}}},
  booktitle  = {2011 {{IEEE}} 26th {{Annual Symposium}} on {{Logic}} in {{Computer Science}}},
  author     = {Birkedal, Lars and Mogelberg, Rasmus Ejlers and Schwinghammer, Jan and Stovring, Kristian},
  year       = 2011,
  month      = jun,
  pages      = {55--64},
  issn       = {1043-6871},
  doi        = {10.1109/LICS.2011.16}
}

@article{bleiVariationalInferenceReview2017,
  title      = {Variational {{Inference}}: {{A Review}} for {{Statisticians}}},
  shorttitle = {Variational {{Inference}}},
  author     = {Blei, David M. and Kucukelbir, Alp and McAuliffe, Jon D.},
  year       = 2017,
  month      = apr,
  journal    = {Journal of the American Statistical Association},
  volume     = {112},
  number     = {518},
  pages      = {859--877},
  publisher  = {Taylor \& Francis},
  issn       = {0162-1459},
  doi        = {10.1080/01621459.2017.1285773}
}

@article{bowersStochasticLazyKnowledge2025,
  title     = {Stochastic {{Lazy Knowledge Compilation}} for {{Inference}} in {{Discrete Probabilistic Programs}}},
  author    = {Bowers, Maddy and Lew, Alexander K. and Tenenbaum, Joshua B. and {Solar-Lezama}, Armando and Mansinghka, Vikash K.},
  year      = 2025,
  month     = jun,
  journal   = {Proceedings of the ACM on Programming Languages},
  volume    = {9},
  number    = {PLDI},
  pages     = {1863--1887},
  publisher = {Association for Computing Machinery (ACM)},
  issn      = {2475-1421},
  doi       = {10.1145/3729325},
  copyright = {http://www.acm.org/publications/policies/copyright\_policy\#Background},
  langid    = {english}
}

@article{chiangExactRecursiveProbabilistic2023,
  title    = {Exact {{Recursive Probabilistic Programming}}},
  author   = {Chiang, David and McDonald, Colin and Shan, Chung-chieh},
  year     = 2023,
  month    = apr,
  journal  = {Proceedings of the ACM on Programming Languages},
  volume   = {7},
  number   = {OOPSLA1},
  pages    = {665--695},
  issn     = {2475-1421},
  doi      = {10.1145/3586050},
  langid   = {english}
}

@inproceedings{claessenQuickCheckLightweightTool2000,
  title      = {{{QuickCheck}}: A Lightweight Tool for Random Testing of {{Haskell}} Programs},
  shorttitle = {{{QuickCheck}}},
  booktitle  = {Proceedings of the Fifth {{ACM SIGPLAN}} International Conference on {{Functional}} Programming},
  author     = {Claessen, Koen and Hughes, John},
  year       = 2000,
  month      = sep,
  series     = {{{ICFP}} '00},
  pages      = {268--279},
  publisher  = {Association for Computing Machinery},
  address    = {New York, NY, USA},
  doi        = {10.1145/351240.351266},
  isbn       = {978-1-58113-202-1}
}

@inproceedings{claretBayesianInferenceUsing2013,
  title     = {Bayesian Inference Using Data Flow Analysis},
  booktitle = {Proceedings of the 2013 9th {{Joint Meeting}} on {{Foundations}} of {{Software Engineering}}},
  author    = {Claret, Guillaume and Rajamani, Sriram K. and Nori, Aditya V. and Gordon, Andrew D. and Borgstr{\"o}m, Johannes},
  year      = 2013,
  month     = aug,
  pages     = {92--102},
  publisher = {ACM},
  address   = {Saint Petersburg Russia},
  doi       = {10.1145/2491411.2491423},
  isbn      = {978-1-4503-2237-9},
  langid    = {english}
}

@article{dashAffineMonadsLazy2023,
  title    = {Affine {{Monads}} and {{Lazy Structures}} for {{Bayesian Programming}}},
  author   = {Dash, Swaraj and Kaddar, Younesse and Paquet, Hugo and Staton, Sam},
  year     = 2023,
  month    = jan,
  journal  = {LazyPPL: Affine monads and lazy structures for Bayesian programming},
  volume   = {7},
  number   = {POPL},
  pages    = {46:1338--46:1368},
  doi      = {10.1145/3571239}
}

@inproceedings{davisWhyArentRegular2019,
  title      = {Why Aren't Regular Expressions a Lingua Franca? An Empirical Study on the Re-Use and Portability of Regular Expressions},
  shorttitle = {Why Aren't Regular Expressions a Lingua Franca?},
  booktitle  = {Proceedings of the 2019 27th {{ACM Joint Meeting}} on {{European Software Engineering Conference}} and {{Symposium}} on the {{Foundations}} of {{Software Engineering}}},
  author     = {Davis, James C. and Michael Iv, Louis G. and Coghlan, Christy A. and Servant, Francisco and Lee, Dongyoon},
  year       = 2019,
  month      = aug,
  pages      = {443--454},
  publisher  = {ACM},
  address    = {Tallinn Estonia},
  doi        = {10.1145/3338906.3338909},
  isbn       = {978-1-4503-5572-8},
  langid     = {english}
}

@inproceedings{dong2020learning,
  title     = {Learning Space Partitions for Nearest Neighbor Search},
  booktitle = {Eighth International Conference on Learning Representations ({{ICLR}})},
  author    = {Dong, Yihe and Indyk, Piotr and Razenshteyn, Ilya and Wagner, Tal},
  year      = 2020,
  month     = apr
}

@inproceedings{dubousquetExpressingImplementingOperational1998,
  title     = {Expressing and Implementing Operational Profiles for Reactive Software Validation},
  booktitle = {Proceedings {{Ninth International Symposium}} on {{Software Reliability Engineering}} ({{Cat}}. {{No}}.{{98TB100257}})},
  author    = {{du Bousquet}, L. and Ouabdesselam, F. and Richier, J.-L.},
  year      = 1998,
  month     = nov,
  pages     = {222--230},
  issn      = {1071-9458},
  doi       = {10.1109/ISSRE.1998.730885}
}

@inproceedings{ellisDreamCoderBootstrappingInductive2021,
  title      = {{{DreamCoder}}: Bootstrapping Inductive Program Synthesis with Wake-Sleep Library Learning},
  shorttitle = {{{DreamCoder}}},
  booktitle  = {Proceedings of the 42nd {{ACM SIGPLAN International Conference}} on {{Programming Language Design}} and {{Implementation}}},
  author     = {Ellis, Kevin and Wong, Catherine and Nye, Maxwell and {Sabl{\'e}-Meyer}, Mathias and Morales, Lucas and Hewitt, Luke and Cary, Luc and {Solar-Lezama}, Armando and Tenenbaum, Joshua B.},
  year       = 2021,
  month      = jun,
  pages      = {835--850},
  publisher  = {ACM},
  address    = {Virtual Canada},
  doi        = {10.1145/3453483.3454080},
  isbn       = {978-1-4503-8391-2},
  langid     = {english}
}

@book{gelmanBayesianDataAnalysis2013,
  title     = {Bayesian {{Data Analysis}}},
  author    = {Gelman, Andrew and Carlin, John B. and Stern, Hal S. and Dunson, David B. and Vehtari, Aki and Rubin, Donald B.},
  year      = 2013,
  month     = nov,
  edition   = {0},
  publisher = {{Chapman and Hall/CRC}},
  doi       = {10.1201/b16018},
  isbn      = {978-0-429-11307-9},
  langid    = {english}
}

@inproceedings{gengGrammarConstrained2023,
  title     = {Grammar-{{Constrained Decoding}} for {{Structured NLP Tasks}} without {{Finetuning}}},
  booktitle = {Proceedings of the 2023 {{Conference}} on {{Empirical Methods}} in {{Natural Language Processing}}},
  author    = {Geng, Saibo and Josifoski, Martin and Peyrard, Maxime and West, Robert},
  editor    = {Bouamor, Houda and Pino, Juan and Bali, Kalika},
  year      = 2023,
  month     = dec,
  pages     = {10932--10952},
  publisher = {Association for Computational Linguistics},
  address   = {Singapore},
  doi       = {10.18653/v1/2023.emnlp-main.674}
}

@article{ghahramaniProbabilisticMachineLearning2015,
  title     = {Probabilistic Machine Learning and Artificial Intelligence},
  author    = {Ghahramani, Zoubin},
  year      = 2015,
  month     = may,
  journal   = {Nature},
  volume    = {521},
  number    = {7553},
  pages     = {452--459},
  publisher = {Nature Publishing Group},
  issn      = {1476-4687},
  doi       = {10.1038/nature14541},
  copyright = {2015 Springer Nature Limited},
  langid    = {english}
}

@inproceedings{gittensExtendedOperationalProfile2004,
  title     = {An Extended Operational Profile Model},
  booktitle = {15th {{International Symposium}} on {{Software Reliability Engineering}}},
  author    = {Gittens, M. and Lutfiyya, H. and Bauer, M.},
  year      = 2004,
  month     = nov,
  pages     = {314--325},
  issn      = {1071-9458},
  doi       = {10.1109/ISSRE.2004.8}
}

@article{goldsteinParsingRandomness2022,
  title    = {Parsing Randomness},
  author   = {Goldstein, Harrison and Pierce, Benjamin C.},
  year     = 2022,
  month    = oct,
  journal  = {Proceedings of the ACM on Programming Languages},
  volume   = {6},
  number   = {OOPSLA2},
  pages    = {89--113},
  issn     = {2475-1421},
  doi      = {10.1145/3563291},
  langid   = {english}
}

@article{goldsteinReflectingRandomGeneration2023,
  title    = {Reflecting on {{Random Generation}}},
  author   = {Goldstein, Harrison and Frohlich, Samantha and Wang, Meng and Pierce, Benjamin C.},
  year     = 2023,
  month    = aug,
  journal  = {Proceedings of the ACM on Programming Languages},
  volume   = {7},
  number   = {ICFP},
  pages    = {322--355},
  issn     = {2475-1421},
  doi      = {10.1145/3607842},
  langid   = {english}
}

@misc{goodmanChurchLanguageGenerative2014,
  title         = {Church: A Language for Generative Models},
  shorttitle    = {Church},
  author        = {Goodman, Noah and Mansinghka, Vikash and Roy, Daniel M. and Bonawitz, Keith and Tenenbaum, Joshua B.},
  year          = 2014,
  month         = jul,
  number        = {arXiv:1206.3255},
  eprint        = {1206.3255},
  primaryclass  = {cs},
  publisher     = {arXiv},
  doi           = {10.48550/arXiv.1206.3255},
  archiveprefix = {arXiv}
}

@inproceedings{gordonProbabilisticProgramming2014,
  title     = {Probabilistic Programming},
  booktitle = {Future of {{Software Engineering Proceedings}}},
  author    = {Gordon, Andrew D. and Henzinger, Thomas A. and Nori, Aditya V. and Rajamani, Sriram K.},
  year      = 2014,
  month     = may,
  series    = {{{FOSE}} 2014},
  pages     = {167--181},
  publisher = {Association for Computing Machinery},
  address   = {New York, NY, USA},
  doi       = {10.1145/2593882.2593900},
  isbn      = {978-1-4503-2865-4}
}

@article{gregersenAlmostSure2024,
  title   = {Almost-{{Sure Termination}} by {{Guarded Refinement}}},
  author  = {Gregersen, Simon Oddershede and Aguirre, Alejandro and Haselwarter, Philipp G. and Tassarotti, Joseph and Birkedal, Lars},
  year    = 2024,
  month   = aug,
  journal = {Proceedings of the ACM on Programming Languages},
  volume  = {8},
  number  = {ICFP},
  pages   = {203--233},
  issn    = {2475-1421},
  doi     = {10.1145/3674632},
  langid  = {english}
}

@inproceedings{hartmannStatisticalAnalysisOperational2016,
  title     = {A {{Statistical Analysis}} of {{Operational Profile Driven Testing}}},
  booktitle = {2016 {{IEEE International Conference}} on {{Software Quality}}, {{Reliability}} and {{Security Companion}} ({{QRS-C}})},
  author    = {Hartmann, Herman},
  year      = 2016,
  month     = aug,
  pages     = {109--116},
  doi       = {10.1109/QRS-C.2016.18}
}

@article{holtzenScalingExactInference2020,
  title    = {Scaling Exact Inference for Discrete Probabilistic Programs},
  author   = {Holtzen, Steven and Van Den Broeck, Guy and Millstein, Todd},
  year     = 2020,
  month    = nov,
  journal  = {Proceedings of the ACM on Programming Languages},
  volume   = {4},
  number   = {OOPSLA},
  pages    = {1--31},
  issn     = {2475-1421},
  doi      = {10.1145/3428208},
  langid   = {english}
}

@misc{hwangInducingProbabilisticPrograms2011,
  title         = {Inducing {{Probabilistic Programs}} by {{Bayesian Program Merging}}},
  author        = {Hwang, Irvin and Stuhlm{\"u}ller, Andreas and Goodman, Noah D.},
  year          = 2011,
  month         = oct,
  number        = {arXiv:1110.5667},
  eprint        = {1110.5667},
  primaryclass  = {cs},
  publisher     = {arXiv},
  doi           = {10.48550/arXiv.1110.5667},
  archiveprefix = {arXiv},
  langid        = {english}
}

@article{johnsonPCFGModelsLinguistic1998,
  title     = {{{PCFG Models}} of {{Linguistic Tree Representations}}},
  author    = {Johnson, Mark},
  editor    = {Hirschberg, Julia},
  year      = 1998,
  journal   = {Computational Linguistics},
  volume    = {24},
  number    = {4},
  pages     = {613--632},
  publisher = {MIT Press},
  address   = {Cambridge, MA}
}

@misc{kingmaAdam2017,
  title         = {Adam: {{A Method}} for {{Stochastic Optimization}}},
  shorttitle    = {Adam},
  author        = {Kingma, Diederik P. and Ba, Jimmy},
  year          = 2017,
  month         = jan,
  number        = {arXiv:1412.6980},
  eprint        = {1412.6980},
  primaryclass  = {cs.LG},
  publisher     = {arXiv},
  doi           = {10.48550/arXiv.1412.6980},
  archiveprefix = {arXiv},
  langid        = {english}
}

@incollection{kiselyovEmbeddedProbabilisticProgramming2009,
  title     = {Embedded {{Probabilistic Programming}}},
  booktitle = {Domain-{{Specific Languages}}},
  author    = {Kiselyov, Oleg and Shan, Chung-chieh},
  editor    = {Taha, Walid Mohamed},
  year      = 2009,
  volume    = {5658},
  pages     = {360--384},
  publisher = {Springer Berlin Heidelberg},
  address   = {Berlin, Heidelberg},
  doi       = {10.1007/978-3-642-03034-5_17},
  isbn      = {978-3-642-03033-8 978-3-642-03034-5},
  langid    = {english}
}

@inproceedings{kollerEffectiveBayesianInference1997,
  title     = {Effective {{Bayesian}} Inference for Stochastic Programs},
  booktitle = {Proceedings of the Fourteenth National Conference on Artificial Intelligence and Ninth Conference on {{Innovative}} Applications of Artificial Intelligence},
  author    = {Koller, Daphne and McAllester, David and Pfeffer, Avi},
  year      = 1997,
  month     = jul,
  series    = {{{AAAI}}'97/{{IAAI}}'97},
  pages     = {740--747},
  publisher = {AAAI Press},
  address   = {Providence, Rhode Island},
  isbn      = {978-0-262-51095-0}
}

@inproceedings{kraskaCaseLearnedIndex2018b,
  title     = {The {{Case}} for {{Learned Index Structures}}},
  booktitle = {Proceedings of the 2018 {{International Conference}} on {{Management}} of {{Data}}},
  author    = {Kraska, Tim and Beutel, Alex and Chi, Ed H. and Dean, Jeffrey and Polyzotis, Neoklis},
  year      = 2018,
  month     = may,
  series    = {{{SIGMOD}} '18},
  pages     = {489--504},
  publisher = {Association for Computing Machinery},
  address   = {New York, NY, USA},
  doi       = {10.1145/3183713.3196909},
  isbn      = {978-1-4503-4703-7}
}

@article{lakeHumanlevelConceptLearning2015,
  title     = {Human-Level Concept Learning through Probabilistic Program Induction},
  author    = {Lake, Brenden M. and Salakhutdinov, Ruslan and Tenenbaum, Joshua B.},
  year      = 2015,
  month     = dec,
  journal   = {Science},
  volume    = {350},
  number    = {6266},
  pages     = {1332--1338},
  issn      = {0036-8075, 1095-9203},
  doi       = {10.1126/science.aab3050},
  copyright = {http://www.sciencemag.org/about/science-licenses-journal-article-reuse},
  langid    = {english}
}

@article{liCompilingProbabilisticPrograms2024,
  title    = {Compiling {{Probabilistic Programs}} for {{Variable Elimination}} with {{Information Flow}}},
  author   = {Li, Jianlin and Wang, Eric and Zhang, Yizhou},
  year     = 2024,
  month    = jun,
  journal  = {Proceedings of the ACM on Programming Languages},
  volume   = {8},
  number   = {PLDI},
  pages    = {1755--1780},
  issn     = {2475-1421},
  doi      = {10.1145/3656448},
  langid   = {english}
}

@article{liulimited1989,
  title    = {On the Limited Memory {{BFGS}} Method for Large Scale Optimization},
  author   = {Liu, Dong C. and Nocedal, Jorge},
  year     = 1989,
  month    = aug,
  journal  = {Mathematical Programming},
  volume   = {45},
  number   = {1},
  pages    = {503--528},
  issn     = {1436-4646},
  doi      = {10.1007/BF01589116},
  langid   = {english}
}

@inproceedings{malyshevGettingMostOut2018,
  title     = {Getting the Most out of Wikidata: {{Semantic}} Technology Usage in Wikipedia's Knowledge Graph},
  booktitle = {Proceedings of the 17th International Semantic Web Conference ({{ISWC}}'18)},
  author    = {Malyshev, Stanislav and Kr{\"o}tzsch, Markus and Gonz{\'a}lez, Larry and Gonsior, Julius and Bielefeldt, Adrian},
  editor    = {Vrande{\v c}i{\'c}, Denny and Bontcheva, Kalina and {Su{\'a}rez-Figueroa}, Mari Carmen and Presutti, Valentina and Celino, Irene and Sabou, Marta and Kaffee, Lucie-Aim{\'e}e and Simperl, Elena},
  year      = 2018,
  series    = {{{LNCS}}},
  volume    = {11137},
  pages     = {376--394},
  publisher = {Springer}
}

@misc{meentIntroductionProbabilisticProgramming2021,
  title         = {An {{Introduction}} to {{Probabilistic Programming}}},
  author        = {van de Meent, Jan-Willem and Paige, Brooks and Yang, Hongseok and Wood, Frank},
  year          = 2021,
  month         = oct,
  number        = {arXiv:1809.10756},
  eprint        = {1809.10756},
  primaryclass  = {stat},
  publisher     = {arXiv},
  doi           = {10.48550/arXiv.1809.10756},
  archiveprefix = {arXiv},
  langid        = {english}
}

@article{muggletonInductiveLogicProgramming1994,
  title      = {Inductive {{Logic Programming}}: {{Theory}} and Methods},
  shorttitle = {Inductive {{Logic Programming}}},
  author     = {Muggleton, Stephen and De Raedt, Luc},
  year       = 1994,
  month      = may,
  journal    = {The Journal of Logic Programming},
  volume     = {19--20},
  pages      = {629--679},
  issn       = {07431066},
  doi        = {10.1016/0743-1066(94)90035-3},
  copyright  = {https://www.elsevier.com/tdm/userlicense/1.0/},
  langid     = {english}
}

@article{musaOperationalProfilesSoftwarereliability1993,
  title    = {Operational Profiles in Software-Reliability Engineering},
  author   = {Musa, J.D.},
  year     = 1993,
  month    = mar,
  journal  = {IEEE Software},
  volume   = {10},
  number   = {2},
  pages    = {14--32},
  issn     = {1937-4194},
  doi      = {10.1109/52.199724}
}

@inproceedings{nakanomodality2000a,
  title     = {A Modality for Recursion},
  booktitle = {Proceedings {{Fifteenth Annual IEEE Symposium}} on {{Logic}} in {{Computer Science}} ({{Cat}}. {{No}}.{{99CB36332}})},
  author    = {Nakano, H.},
  year      = 2000,
  month     = jun,
  pages     = {255--266},
  issn      = {1043-6871},
  doi       = {10.1109/LICS.2000.855774}
}

@inproceedings{narayanan2016probabilistic,
  title        = {Probabilistic inference by program transformation in Hakaru (system description)},
  author       = {Narayanan, Praveen and Carette, Jacques and Romano, Wren and Shan, Chung{-}chieh and Zinkov, Robert},
  booktitle    = {International Symposium on Functional and Logic Programming - 13th International Symposium, {FLOPS} 2016, Kochi, Japan, March 4-6, 2016, Proceedings},
  pages        = {62--79},
  year         = {2016},
  organization = {Springer},
  url          = {http://dx.doi.org/10.1007/978-3-319-29604-3_5},
  doi          = {10.1007/978-3-319-29604-3_5}
}

@inproceedings{noriEfficientSynthesisProbabilistic2015,
  title     = {Efficient Synthesis of Probabilistic Programs},
  booktitle = {Proceedings of the 36th {{ACM SIGPLAN Conference}} on {{Programming Language Design}} and {{Implementation}}},
  author    = {Nori, Aditya V. and Ozair, Sherjil and Rajamani, Sriram K. and Vijaykeerthy, Deepak},
  year      = 2015,
  month     = jun,
  pages     = {208--217},
  publisher = {ACM},
  address   = {Portland OR USA},
  doi       = {10.1145/2737924.2737982},
  isbn      = {978-1-4503-3468-6},
  langid    = {english}
}

@article{papenbrockFunctionalDependencyDiscovery2015,
  title      = {Functional Dependency Discovery: {{An}} Experimental Evaluation of Seven Algorithms},
  shorttitle = {Functional Dependency Discovery},
  author     = {Papenbrock, Thorsten and Ehrlich, Jens and Marten, Jannik and Neubert, Tommy and Rudolph, Jan-Peer and Sch{\"o}nberg, Martin and Zwiener, Jakob and Naumann, Felix},
  year       = 2015,
  journal    = {Proceedings of the VLDB Endowment},
  volume     = {8},
  number     = {10},
  pages      = {1082--1093},
  publisher  = {VLDB Endowment}
}

@misc{paveseInputs2018,
  title         = {Inputs from {{Hell}}: {{Generating Uncommon Inputs}} from {{Common Samples}}},
  shorttitle    = {Inputs from {{Hell}}},
  author        = {Pavese, Esteban and Soremekun, Ezekiel and Havrikov, Nikolas and Grunske, Lars and Zeller, Andreas},
  year          = 2018,
  month         = dec,
  number        = {arXiv:1812.07525},
  eprint        = {1812.07525},
  primaryclass  = {cs.SE},
  publisher     = {arXiv},
  doi           = {10.48550/arXiv.1812.07525},
  archiveprefix = {arXiv},
  langid        = {english}
}

@article{Pfeffer2015LazyFI,
  title   = {Lazy Factored Inference for Functional Probabilistic Programming},
  author  = {Avi Pfeffer and Brian E. Ruttenberg and Amy Lynn Sliva and Michael Howard and Glenn Takata},
  journal = {ArXiv},
  year    = {2015},
  volume  = {abs/1509.03564},
  url     = {https://api.semanticscholar.org/CorpusID:1257469}
}

@article{rothHardnessApproximateReasoning1996,
  title     = {On the Hardness of Approximate Reasoning},
  author    = {Roth, Dan},
  year      = 1996,
  month     = apr,
  journal   = {Artificial Intelligence},
  volume    = {82},
  number    = {1-2},
  pages     = {273--302},
  issn      = {00043702},
  doi       = {10.1016/0004-3702(94)00092-1},
  copyright = {https://www.elsevier.com/tdm/userlicense/1.0/},
  langid    = {english}
}

@article{saadBayesianSynthesisProbabilistic2019,
  title         = {Bayesian {{Synthesis}} of {{Probabilistic Programs}} for {{Automatic Data Modeling}}},
  author        = {Saad, Feras A. and {Cusumano-Towner}, Marco F. and Schaechtle, Ulrich and Rinard, Martin C. and Mansinghka, Vikash K.},
  year          = 2019,
  month         = jan,
  journal       = {Proceedings of the ACM on Programming Languages},
  volume        = {3},
  number        = {POPL},
  eprint        = {1907.06249},
  primaryclass  = {cs},
  pages         = {1--32},
  issn          = {2475-1421},
  doi           = {10.1145/3290350},
  archiveprefix = {arXiv},
  langid        = {english}
}

@article{scibiorFunctional2018,
  title   = {Functional Programming for Modular {{Bayesian}} Inference},
  author  = {{\'S}cibior, Adam and Kammar, Ohad and Ghahramani, Zoubin},
  year    = 2018,
  month   = jul,
  journal = {Proceedings of the ACM on Programming Languages},
  volume  = {2},
  number  = {ICFP},
  pages   = {1--29},
  issn    = {2475-1421},
  doi     = {10.1145/3236778},
  langid  = {english}
}

@article{smidtsSoftwareTestingOperational2014,
  title      = {Software Testing with an Operational Profile: {{OP}} Definition},
  shorttitle = {Software Testing with an Operational Profile},
  author     = {Smidts, Carol and Mutha, Chetan and Rodr{\'i}guez, Manuel and Gerber, Matthew J.},
  year       = 2014,
  month      = feb,
  journal    = {ACM Comput. Surv.},
  volume     = {46},
  number     = {3},
  pages      = {39:1--39:39},
  issn       = {0360-0300},
  doi        = {10.1145/2518106}
}

@article{stassenModelling2025a,
  title   = {Modelling {{Recursion}} and {{Probabilistic Choice}} in {{Guarded Type Theory}}},
  author  = {Stassen, Philipp and M{\o}gelberg, Rasmus Ejlers and Zwart, Maaike Annebet and Aguirre, Alejandro and Birkedal, Lars},
  year    = 2025,
  month   = jan,
  journal = {Proceedings of the ACM on Programming Languages},
  volume  = {9},
  number  = {POPL},
  pages   = {1417--1445},
  issn    = {2475-1421},
  doi     = {10.1145/3704884},
  langid  = {english}
}

@inproceedings{stolckeInducingProbabilisticGrammars1994,
  title     = {Inducing Probabilistic Grammars by {{Bayesian}} Model Merging},
  booktitle = {Grammatical {{Inference}} and {{Applications}}},
  author    = {Stolcke, Andreas and Omohundro, Stephen},
  editor    = {Carrasco, Rafael C. and Oncina, Jose},
  year      = 1994,
  pages     = {106--118},
  publisher = {Springer},
  address   = {Berlin, Heidelberg},
  doi       = {10.1007/3-540-58473-0_141},
  isbn      = {978-3-540-48985-6},
  langid    = {english}
}

@misc{stuhlmullerDynamic2012,
  title         = {A {{Dynamic Programming Algorithm}} for {{Inference}} in {{Recursive Probabilistic Programs}}},
  author        = {Stuhlm{\"u}ller, Andreas and Goodman, Noah D.},
  year          = 2012,
  month         = sep,
  number        = {arXiv:1206.3555},
  eprint        = {1206.3555},
  primaryclass  = {cs.AI},
  publisher     = {arXiv},
  doi           = {10.48550/arXiv.1206.3555},
  archiveprefix = {arXiv}
}

@inproceedings{tazwarTabVAENovelVAE2024,
  title      = {Tab-{{VAE}}: {{A Novel VAE}} for {{Generating Synthetic Tabular Data}}:},
  shorttitle = {Tab-{{VAE}}},
  booktitle  = {Proceedings of the 13th {{International Conference}} on {{Pattern Recognition Applications}} and {{Methods}}},
  author     = {Tazwar, Syed and Knobbout, Max and Quesada, Enrique and Popa, Mirela},
  year       = 2024,
  pages      = {17--26},
  publisher  = {{SCITEPRESS - Science and Technology Publications}},
  address    = {Rome, Italy},
  doi        = {10.5220/0012302400003654},
  isbn       = {978-989-758-684-2},
  langid     = {english}
}

@article{tjoaTuningRandomGenerators2025,
  title      = {Tuning {{Random Generators}}: {{Property-Based Testing}} as {{Probabilistic Programming}}},
  shorttitle = {Tuning {{Random Generators}}},
  author     = {Tjoa, Ryan and Garg, Poorva and Goldstein, Harrison and Millstein, Todd and Pierce, Benjamin C. and Van Den Broeck, Guy},
  year       = 2025,
  month      = oct,
  journal    = {Proceedings of the ACM on Programming Languages},
  volume     = {9},
  number     = {OOPSLA2},
  pages      = {894--921},
  issn       = {2475-1421},
  doi        = {10.1145/3763082},
  langid     = {english}
}

@inproceedings{vaswaniAttention2017,
  title     = {Attention Is All You Need},
  booktitle = {Advances in Neural Information Processing Systems},
  author    = {Vaswani, Ashish and Shazeer, Noam and Parmar, Niki and Uszkoreit, Jakob and Jones, Llion and Gomez, Aidan N and Kaiser, {\L}ukasz and Polosukhin, Illia},
  editor    = {Guyon, I. and Luxburg, U. Von and Bengio, S. and Wallach, H. and Fergus, R. and Vishwanathan, S. and Garnett, R.},
  year      = 2017,
  volume    = {30},
  publisher = {Curran Associates, Inc.}
}

@article{werbosBackpropagationTimeWhat1990,
  title      = {Backpropagation through Time: What It Does and How to Do It},
  shorttitle = {Backpropagation through Time},
  author     = {Werbos, P.J.},
  year       = 1990,
  month      = oct,
  journal    = {Proceedings of the IEEE},
  volume     = {78},
  number     = {10},
  pages      = {1550--1560},
  issn       = {1558-2256},
  doi        = {10.1109/5.58337}
}

@incollection{xuModelingTabularData2019,
  title     = {Modeling Tabular Data Using Conditional {{GAN}}},
  booktitle = {Proceedings of the 33rd {{International Conference}} on {{Neural Information Processing Systems}}},
  author    = {Xu, Lei and Skoularidou, Maria and {Cuesta-Infante}, Alfredo and Veeramachaneni, Kalyan},
  year      = 2019,
  month     = dec,
  number    = {659},
  pages     = {7335--7345},
  publisher = {Curran Associates Inc.},
  address   = {Red Hook, NY, USA}
}

@article{zekiciTestingSoftwareOperational2000,
  title     = {Testing of Software with an Operational Profile},
  author    = {{\"O}zekici, S{\"u}leyman {\"O}zekici and Alt{\'i}nel, {\.I}. Kuban and {\"O}z{\c c}eliky{\"u}rek, Selma},
  year      = 2000,
  month     = dec,
  journal   = {Naval Research Logistics},
  volume    = {47},
  number    = {8},
  pages     = {620--634},
  issn      = {0894-069X, 1520-6750},
  doi       = {10.1002/1520-6750(200012)47:8<620::AID-NAV2>3.0.CO;2-V},
  copyright = {http://doi.wiley.com/10.1002/tdm\_license\_1.1},
  langid    = {english}
}

@book{pierceAdvancedTopicsTypes2005,
  title     = {Advanced {{Topics}} in {{Types}} and {{Programming Languages}}},
  editor    = {Pierce, Benjamin C.},
  year      = 2005,
  publisher = {MIT Press},
  address   = {Cambridge, Mass},
  isbn      = {978-0-262-16228-9},
  lccn      = {QA76.7 .A36 2005}
}

\newpage

\appendix

\section{Probabilistic semantics}
\label{apx:probSemantics}

\subsection{Proof of \autoref{lem:probDist}}
We prove the result by structural induction on the denotation.
For base values, variable references, and sampling operations, the denotations are point masses.
A constructor application matches only the correct constructor and distributes its mass pointwise across its arguments.
Binding forms such as let-bindings and function applications marginalize the distributions of their arguments over the bound variables.
The equation for case destructuring is the only one with an additional case: its final summand $\muDenot{e}(\bot)$ is the only term (apart from the function-context approximants described below) that places mass on $\bot$.

\subsection{Function fixed points and approximants }

The zeroth approximant of the function context $\phi_0$ is a point mass on $\bot$, rather than the zero distribution $\phi_0(f)(\overline{\delta})(\delta') = 0$.
This is the appropriate choice in this context.
To see why, consider the $\fn{gen}()$ example on page~\pageref{ex:gen}.
The zero distribution would assign zero probability to every finite list at every approximant $\phi_k$, even though $\phi_k$ represents $\fn{gen}$ unrolled $k$ times.
Starting at $\bot$ instead allows $\phi_k$ to move its probability mass to lists of length $k$, as intended.

\subsection{Proof sketch of \autoref{thm:refinement}}
We build the coupling by structural induction on $\muDenot{\cdot}$.
Monotonicity in $\rho$ is needed to carry the induction through $\kw{case}$, because the environment is extended with variables bound to the scrutinee's children, and the refined value of a child may be above its coarse value.

\subsection{Proof of \autoref{thm:correctness}}
\label{apx:soundnessProof}

The proof has two parts.
The first is a semantic reading of the modal index: an expression typed at $\later^n A$, evaluated under the $k$-th approximant, produces partial values whose undefined parts all lie at depth at least $k - n$, up to a constant that accounts for the finitely many non-recursive call layers.
The second observes that once an approximant is defined below the height of a complete observation, no further unrolling can move mass onto or off of that observation.

\paragraph{Notation.}
Throughout, fix a well-typed program
\[
    P = \defin f_1(\overline{x_1 : \tau_1}) : \tau_1' \coloneq e_1;\;\dots\;%
        \defin f_m(\overline{x_m : \tau_m}) : \tau_m' \coloneq e_m;\;%
        e_{m+1},
\]
where we write $e_{m+1}$ for the main body, and let $\Sigma$ be the signature accumulated by \textsc{T-Defn}.
By \textsc{T-Defn}, the body $e_i$ of $f_i$ is typed against a signature containing exactly $f_1, \dots, f_i$, so every call in $e_i$ is to some $f_j$ with $j \leq i$, and the call is a self-call exactly when $j = i$.
We say that $e_i$ has \emph{position} $i$, and that the main body has position $m + 1$.

For a partial value $\delta \in \calV_\bot$, define its \emph{definedness depth} $\bdepth{\delta} \in \N \cup \{\infty\}$ as the depth of its shallowest $\bot$, or $\infty$ if it has none:
\[
    \bdepth{\bot} = 0, \qquad
    \bdepth{C(\delta_1, \dots, \delta_r)} = 1 + \min_{i \leq r} \bdepth{\delta_i},
\]
with the convention that the minimum over an empty sequence is $\infty$. Thus, nullary constructors and, inductively, all complete values have infinite definedness depth.
For a complete value $v \in \calV$, define its \emph{height} $\hgt{v}$ by $\hgt{C} = 1$ and $\hgt{C(v_1, \dots, v_r)} = 1 + \max_i \hgt{v_i}$.
Both functions are computable.
We will need the following elementary relationship between them.

\begin{lemma}[Defined below the observation]
    \label{lem:definedBelow}
    Let $v \in \calV$ and $\delta \in \calV_\bot$ with $\delta \sqsubseteq v$.
    If $\bdepth{\delta} \geq \hgt{v}$, then $\delta = v$.
\end{lemma}

\begin{proof}
    By induction on $v$.
    Since $\hgt{v} \geq 1$, $\delta \neq \bot$, so $\delta = C(\overline{\delta})$ and $v = C(\overline{v})$ for the same constructor with $\delta_i \sqsubseteq v_i$ for each $i$.
    Then $\bdepth{\delta_i} \geq \bdepth{\delta} - 1 \geq \hgt{v} - 1 \geq \hgt{v_i}$, so $\delta_i = v_i$ by the induction hypothesis.
\end{proof}

\paragraph{Levels.}
The modal index is a \emph{relative} quantity, and the semantics of \autoref{fig:semantics} charges one approximant per call, whether or not the call is recursive.
To relate these quantities, we introduce a level $\ell \in \Z$ that decreases with each call.
Because the non-self call graph is acyclic and ordered by position, a body at position $i$ evaluated under $\phi_k$ has at most $i - 1$ non-recursive calls stacked beneath it. We will show that the level $\ell = k - i + 1$ is always available to it.
Negative levels are permitted and make every claim below vacuous, which is what allows the induction to start at $\phi_0$.

For a context $\Gamma$ and an environment $\rho$ defined on $\dom(\Gamma)$, write $\rho \satL{\ell} \Gamma$ when $\bdepth{\rho(x)} \geq \ell - n$ for every $x : \later^n A \in \Gamma$.

\begin{lemma}[Semantic typing]
    \label{lem:semanticTyping}
    Let $\judge[\Sigma; \Gamma; f]{e}{\later^n A}$ be derivable, where $e$ occurs in the body at position $i$ (so $f = f_i$, or $f = \cdot$ and $i = m + 1$).
    Then for every $k \geq 0$, every level $\ell \leq k - i + 1$, and every $\rho \satL{\ell} \Gamma$, every $\delta$ in the support of $\muDenot[{\rho,\phi_k}]{e}$ satisfies $\bdepth{\delta} \geq \ell - n$.
\end{lemma}

\begin{proof}
    By induction on $k$, and for each $k$ by an inner induction on the typing derivation.
    We use the outer induction hypothesis only for calls, whose callee bodies are evaluated under $\phi_{k-1}$. For all other subterms, which are evaluated under $\phi_k$, we use the inner induction hypothesis.
    Fix $k$, $\ell \leq k - i + 1$, and $\rho \satL{\ell} \Gamma$.

    \emph{\textsc{T-Var}.}
    The support is $\{\rho(x)\}$, and $\bdepth{\rho(x)} \geq \ell - n$ by $\rho \satL{\ell} \Gamma$.

    \emph{\textsc{T-Flip}, \textsc{T-Geom}, \textsc{T-Constr-Null}.}
    The support consists of complete values, whose definedness depth is $\infty$.

    \emph{\textsc{T-Later-Sub}.}
    The inner hypothesis gives $\bdepth{\delta} \geq \ell - m \geq \ell - n$ since $m \leq n$.

    \emph{\textsc{T-Constr}.}
    The support of $\muDenot[{\rho,\phi_k}]{C(\overline{e})}$ consists of values $C(\overline{\delta})$ with each $\delta_r$ in the support of $\muDenot[{\rho,\phi_k}]{e_r}$.
    Each $e_r$ is typed at $\later^{n+1} A_r$, so $\bdepth{\delta_r} \geq \ell - n - 1$ by the inner hypothesis, and hence $\bdepth{C(\overline{\delta})} = 1 + \min_r \bdepth{\delta_r} \geq \ell - n$.
    This is the step that converts one layer of observable output into one unit of modal debt.

    \emph{\textsc{T-Let}.}
    Mass on $\delta$ arises from some $\delta'$ in the support of $\muDenot[{\rho,\phi_k}]{e}$ together with $\delta$ in the support of $\muDenot[{\rho[x \mapsto \delta'],\phi_k}]{e'}$.
    If $e$ is typed at $\later^{n_1} A_1$ then $\bdepth{\delta'} \geq \ell - n_1$ by the inner hypothesis, so $\rho[x \mapsto \delta'] \satL{\ell} \Gamma, x : \later^{n_1} A_1$, and the inner hypothesis for $e'$ gives $\bdepth{\delta} \geq \ell - n$.

    \emph{\textsc{T-Case}.}
    Let the scrutinee $e$ be typed at $\later^{n_0} A$ and the branches at $\later^n A'$ with $n_0 \leq n$.
    By the equation for $\kw{case}$ in \autoref{fig:semantics}, mass reaches $\delta$ in one of two ways.
    Either the scrutinee produces $C_i(\overline{\delta'})$ and branch $i$ produces $\delta$ under $\rho[\overline{x_i} \mapsto \overline{\delta'}]$; then $\bdepth{C_i(\overline{\delta'})} \geq \ell - n_0$ by the inner hypothesis, so each $\bdepth{\delta'_r} \geq \ell - n_0 - 1$, the extended environment satisfies $\Gamma, \overline{x_i : \later^{n_0 + 1} A_{i}}$ at level $\ell$, and the inner hypothesis for the branch gives $\bdepth{\delta} \geq \ell - n$.
    Or the scrutinee produces $\bot$ and $\delta = \bot$; this requires $0 = \bdepth{\bot} \geq \ell - n_0$, so $\ell \leq n_0 \leq n$, and the claim $\bdepth{\bot} = 0 \geq \ell - n$ holds.
    This second case is the only place where we use the side condition $n_0 \leq n$: a branch may not promise to be more defined than a scrutinee that may still be undefined.

    \emph{\textsc{T-Call}.}
    Let the callee be $g = f_j$ with $j \leq i$ and $\Sigma(f_j) = (\later^{m_1} A_1, \dots, \later^{m_r} A_r) \to \later^{n_j} A'$, and let $s$ be the shift chosen in the rule, so that each argument $e_t$ is typed at $\later^{m_t + s} A_t$ and the call at $\later^{n_j + s} A'$.
    By the inner hypothesis, every $\delta'_t$ in the support of $\muDenot[{\rho,\phi_k}]{e_t}$ has $\bdepth{\delta'_t} \geq \ell - m_t - s = (\ell - s) - m_t$; that is, the argument environment $\rho_j = \{x_{jt} \mapsto \delta'_t\}$ satisfies $\rho_j \satL{\ell - s} \overline{x_j : \tau_j}$.

    If $k = 0$, then $\phi_0(f_j)(\overline{\delta'})$ is the point mass on $\bot$, and $\bdepth{\bot} = 0 \geq \ell - n_j - s$ holds because $\ell \leq k - i + 1 = 1 - i \leq 0$.

    If $k \geq 1$, then $\phi_k(f_j)(\overline{\delta'}) = \muDenot[{\rho_j,\phi_{k-1}}]{e_j}$, the body of $f_j$ at position $j$ evaluated under $\phi_{k-1}$.
    We apply the outer induction hypothesis at $k - 1$ to the derivation $\judge[\Sigma;\, \overline{x_j : \tau_j};\, f_j]{e_j}{\later^{n_j} A'}$ supplied by \textsc{T-Defn}, at level $\ell - s$.
    This requires $\ell - s \leq (k - 1) - j + 1 = k - j$.
    If $j < i$ (a non-recursive call), then $k - j \geq k - i + 1 \geq \ell \geq \ell - s$ since $s \geq 0$.
    If $j = i$ (a self-call), the rule requires $s \geq 1$, so $\ell - s \leq \ell - 1 \leq k - i = k - j$.
    In either case the hypothesis applies and yields $\bdepth{\delta} \geq (\ell - s) - n_j$ for every $\delta$ in the support of $\phi_k(f_j)(\overline{\delta'})$, which is the claim at $\later^{n_j + s} A'$.
    Here the acyclicity of the non-self call graph is what pays for the $i - 1$ slack in the level: a non-recursive callee sits at a strictly smaller position and so has one more approximant to spend, whereas a self-call must pay for its own approximant through $s \geq 1$.
\end{proof}

The quantification over all levels $\ell \leq k - i + 1$, rather than only the largest level, is what makes the \textsc{T-Call} case work for an arbitrary shift $s$: the callee's body is invoked at the lower level $\ell - s$, and the lemma must already hold there.
This is the semantic counterpart of \Cref{lem:shift}.

\paragraph{Definedness bound.}
Applying \Cref{lem:semanticTyping} to the premise of \textsc{T-Prog-Main}, with position $i = m + 1$, $\rho = \varnothing$, and largest admissible level $\ell = k - m$, yields:

\begin{corollary}[Definedness of approximants]
    \label{cor:definedness}
    If $\judge[\cdot]{P}{\later^n A}$, then for every $k$, every $\delta$ in the support of $\muDenot[{\varnothing,\phi_k}]{e_{m+1}}$ satisfies $\bdepth{\delta} \geq k - m - n$.
\end{corollary}

In particular, taking $k \geq m + n + 1$ shows that $\muDenot[{\varnothing,\phi_k}]{e_{m+1}}(\bot) = 0$, which is \Cref{lem:whnf} for the main body; the same corollary applied to the body of each $f_i$ at its own position gives \Cref{lem:whnf} for function bodies.

\paragraph{Stabilization.}
It remains to show that once an approximant is defined below the height of an observation, all later approximants agree with it on that observation.

\begin{lemma}[Stabilization]
    \label{lem:stabilization}
    Let $v \in \calV$ and $k \leq k'$, and suppose every $\delta$ in the support of $\muDenot[{\varnothing,\phi_k}]{e_{m+1}}$ satisfies $\bdepth{\delta} \geq \hgt{v}$.
    Then $\dbrack{P}_k(v) = \dbrack{P}_{k'}(v)$.
\end{lemma}

\begin{proof}
    Write $\mu_k$ and $\mu_{k'}$ for the two measures.
    Since $v$ is complete, $\upclos{v} = \{v\}$, so $\dbrack{P}_k(v) = \mu_k(v)$ and likewise for $k'$.
    By \Cref{thm:refinement} there is a coupling $\gamma$ of $\mu_k$ and $\mu_{k'}$ supported on pairs $(\delta, \delta')$ with $\delta \sqsubseteq \delta'$.
    Then
    \[
        \mu_{k'}(v) = \sum_{\delta} \gamma(\delta, v) = \sum_{\delta \sqsubseteq v,\; \mu_k(\delta) > 0} \gamma(\delta, v).
    \]
    Every $\delta$ in this sum lies in the support of $\mu_k$ and satisfies $\delta \sqsubseteq v$, so by hypothesis and \Cref{lem:definedBelow}, $\delta = v$.
    Hence $\mu_{k'}(v) = \gamma(v, v) \leq \mu_k(v)$.
    The reverse inequality $\mu_k(v) \leq \mu_{k'}(v)$ is the monotonicity statement of \Cref{thm:refinement}.
\end{proof}

\paragraph{Conclusion.}
Let $\judge[\cdot]{P}{\later^n A}$ and define
\[
    \beta_P(v) = \hgt{v} + m + n,
\]
which is total and computable.
For $k \geq \beta_P(v)$, \Cref{cor:definedness} gives $\bdepth{\delta} \geq k - m - n \geq \hgt{v}$ for every value in the support of the $k$-th approximant. Thus, by \Cref{lem:stabilization}, $\dbrack{P}_k(v) = \dbrack{P}_{k'}(v)$ for every $k' \geq k$.
Since $\dbrack{P}(v) = \sup_{k'} \dbrack{P}_{k'}(v)$ by definition, and the sequence is constant from $k$ onward, $\dbrack{P}_k(v) = \dbrack{P}(v)$.
This is \Cref{def:obsBoundComp}, and $P$ is OBC.

\paragraph{Remarks.}
The bound $\beta_P$ is linear in the height of the observation with unit slope: one approximant per layer, plus a constant.
The constant $m$ is an artifact of charging non-recursive calls an approximant each in \autoref{fig:semantics}; defining $\Phi$ to inline non-recursive callees would remove it, at the cost of a more complicated functional.
The constant $n$ is the modal index of the main body, and is what permits main bodies such as $\fn{head}(\fn{gen}())$, whose first observable constructor is only available after the generator has been unrolled once.
Nothing in the argument depends on \Cref{thm:finiteDemand}; finiteness of demand is what makes the inference procedure built on \Cref{thm:finiteDemand} realizable as a circuit, but exactness at a finite approximant follows from definedness alone.

\section{Probabilistic inference for \tali programs}
\label{apx:probInference}
A well-typed \tali program has all of its recursive calls guarded by at least one constructor.
One consequence is that every well-typed program eventually reduces to weak-head normal form (WHNF) with probability 1.
In our semantics, this is represented as membership in a set of partial values ${\upclos{C(\bot, \dots, \bot)}}$.

\begin{lemma}[Weak-head normal form]
    \label{lem:whnf}
    Let $e$ be an expression, let $\Sigma$ be a well-formed function signature, and let $\Gamma$ be a variable context defined on the free variables of $e$.
    Let $\rho$ be a variable environment defined on the same variables as $\Gamma$.
    If $\judge[\Sigma; \Gamma; f]{e}{\later^n A}$, then \[
        \lim_{k \to \infty} \muDenot[{\rho,\phi_k}]{e}(\bot) = 0,
    \]
    or equivalently,
    \[
        \sum_{C \in \calC_A} \dbrack{e}_{\rho}(C(\bot, \dots, \bot)) = 1.
    \]
\end{lemma}

This holds even when $e$ itself does not terminate.
As mentioned in \autoref{subsec:fixedPoints}, probability mass can reach $\bot$ in two ways: by case-destructuring $\bot$ or by failing to produce a constructor.
By \textsc{T-Case}, destructuring an unguarded self-call (i.e., a value at a modal type with index $> 0$) is allowed, but incurs a modal ``debt'' that can be repaid only by applying a constructor.
Furthermore, because the function call graph excluding self-calls is acyclic, the number of steps needed to produce a constructor is bounded. Thus, some $k$ exists such that the $(k+1)$-th unrolling produces a constructor.
Consequently, we can marginalize over certain nonterminating executions, provided they place zero mass on $\bot$.

\begin{lemma}[Irrelevance]
    \label{lem:irrelevance}
    Let $l \in [0, 1]$, let $\rho$ be a variable context, and let $\phi$ be a function context (either an approximant or the exact context).
    Let $e_1$ and $e_2$ be well-typed expressions.
    Suppose $\dbrack{e_2}_{\rho[x \mapsto \delta'],\phi}(\delta) = l$ for every partial value $\delta' \in \calV_\bot$.
    Then
    \[
        \dbrack{\letin{x}{e_1}{e_2}}_{\rho,\phi}(\delta) = \dbrack{e_2}_{\rho[x \mapsto \bot],\phi}(\delta)
    \]
\end{lemma}

The proof is immediate: the constant $l$ factors out of the marginalization, leaving the total mass $\sum_{\delta' \in \calV_\bot}\muDenot[\rho,\phi]{e_1}(\delta')$, which is 1 by \autoref{lem:probDist}.
This result also generalizes to other binding forms, including function application and case destructuring.
There is no implicit conditioning on termination here.
$\dbrack{P}(v)$ is an unconditional probability over $\calV_\bot$, and the mass which does not reach $\calV$ is not normalized away.
(In any event, normalization is not a particularly attractive option, as the normalizing constant $\sum_{v \in \calV} \dbrack{P}(v)$ is not computable in general.)

An important consequence is that expressions can discard such unbounded subcomputations without affecting the likelihood of the containing value.
In other words, we can marginalize over discarded subcomputations ``for free,'' since their results are never forced or observed.
This is exactly the motivation for using lazy evaluation in prior literature \citep{Pfeffer2015LazyFI,bowersStochasticLazyKnowledge2025,dashAffineMonadsLazy2023,narayanan2016probabilistic,kiselyovEmbeddedProbabilisticProgramming2009,scibiorFunctional2018}.
This allows us to compute the likelihood of a finite observation from a program such as $\fn{take}(3, \fn{gen}())$, even though $\fn{gen}$ places zero mass on complete ($\bot$-free) finite lists.
Because \autoref{lem:irrelevance} holds for all approximants, it is also useful for bounded unrolling: beyond a certain bound, we need not unroll the infinite tails of $\fn{gen}()$; we need only meet the observation's demand.

Together, the OBC property and the type system that enforces it provide a straightforward knowledge-compilation mechanism for productive recursive programs under our semantics.
The first step toward an inference algorithm is to show that the likelihood at each step along an observation is a sum over a finite set of demands.
\begin{theorem}[Finite demand]
    \label{thm:finiteDemand}
    Given an expression $e$, a variable context $\rho$ containing all free variables in $e$, an approximant function context $\phi_k$, a variable $x$, and a demand $\delta$, define the function \[
        p_{\delta}(d) = \dbrack{e}_{\rho[x \mapsto d],\phi_k}(\delta).
    \]
    Then, for each $\delta' \in \calV_\bot$, there are finitely many demands $\delta_1, \dots, \delta_m$
        and weights $w_1, \dots w_m$ such that \[p_\delta(\delta') = \sum_j w_j \cdot [\delta_j \sqsubseteq \delta'].\]
\end{theorem}

The demands $\delta_j$ are pairwise incompatible, meaning that no pair shares an upper bound; therefore, their upward closures are pairwise disjoint.
The only elimination form in \tali is the $\kw{case}$ expression, and by \textsc{T-Case}, each $\kw{case}$ expression exhaustively spans a finite set of distinct constructors.
Thus the demand placed upon a scrutinee $e$ is generated by a finite decision tree, in which each internal node forces one layer of $e$ and branches on the forced constructor.
Two leaves then differ in the constructor at the first position at which their paths diverge.
Demands which are merely \emph{incomparable}, but not incompatible, cannot result from this forcing.
For example, two such demands might be $\con{Cons}(\bot, \con{Nil})$ and $\con{Cons}(0, \bot)$, which are pointwise incomparable (the orders on the two subterms are opposites of one another), but nonetheless share an upper bound $\con{Cons}(0, \con{Nil})$.
The finite constructor domain of scrutinized values then bounds the tree's branching.

Likewise, restricting the function context to an approximant $\phi_k$ is necessary, not merely cosmetic.
Under $\phi_k$, recursion is unrolled $k$ times, which bounds the depth of the decision tree.
If instead we used the fixed point $\phi$, a function such as
\[
    \defin \fn{consume}(xs:\text{List}) : \text{Unit} \coloneq\\
    \kw{case} xs\; \kw{of} \{\con{Cons}(hd, tl) \to \fn{consume}(tl)\}\\
    \{\con{Nil} \to \con{Unit}\}\\
\]
would place infinite demand $\{\con{Nil}, \con{Cons}(\bot, \con{Nil}), \con{Cons}(\bot, \con{Cons}(\bot, \con{Nil}))\dots\}$ upon its argument $xs$ and the decision tree would have infinite height.
(In any case, such functions are not well-typed due to their unguarded recursion, and \autoref{thm:correctness} can be interpreted as a stabilization guarantee on the demand of a function argument as $k$ grows for well-typed programs.%
\footnote{
    While the example is rejected by the type system, the generality of $\phi_k$ is still needed to avoid this infinite demand; the inference procedure below builds its circuit at a finite approximant, and the formulation given here is what guarantees that circuit is finite.
})

For binding forms such as $\kw{let}$, the equations become finite sums when expressed in terms of demands rather than point masses.
For example,
\[
    \dbrack{\letin{x}{e_1}{e_2}}_{\rho,\phi_k}(\delta) = \sum_{j = 1}^m c_j \cdot \dbrack{e_1}_{\rho,\phi_k}(\delta_j)%
        + c_0 \left(1 - \sum_{j = 1}^m \dbrack{e_1}_{\rho,\phi_k}(\delta_j)\right),
\]

where $m$, $c_0$, $c_1, \dots, c_m$, and $\delta_1, \dots, \delta_m$ are obtained by applying \autoref{thm:finiteDemand} to $e_2$.
Each $\delta_j$ is a demand placed on $x$ by $e_2$.
The same construction reduces all sums over $\calV_\bot$ in \autoref{fig:semantics} to finite sums over demands, yielding the probabilistic inference procedure.
Given a program $P$ and observation $v \in \calV$, we compute the likelihood $\dbrack{P}(v)$ by propagating demands ``outside-in,'' forcing each subcomputation only as much as needed to evaluate each step's likelihood.
We implement this procedure by incrementally building an arithmetic circuit for the likelihood of the demand at each forced step. The finite set of demands and bounded depth of function unrolling ensure that this circuit is finite.

\section{Type rules for the \tali language}
\label{apx:typeRules}

\begin{figure}[t]
    \begin{mathpar}
        \inferrule[T-Var]{
            \Gamma(x) = \tau
        }{
            \judge[\Sigma; \Gamma; f]{x}{\tau}
        }\and
        \inferrule[T-Flip]{
            ~
        }{
            \judge[\Sigma; \Gamma; f]{\flip p}{\mathsf{Bool}}
        }\and
        \inferrule[T-Geom]{
            ~
        }{
            \judge[\Sigma; \Gamma; f]{\geom p}{\mathsf{Int}}
        }\and
        \inferrule[T-Later-Sub]{
            \judge[\Sigma; \Gamma; f]{e}{\later^m A}\\
            m \leq n
        }{
            \judge[\Sigma; \Gamma; f]{e}{\later^n A}
        }\and
        \inferrule[T-Constr-Null]{
            C \in \calC_A \\
            |\sig(C)| = 0
        }{
            \judge[\Sigma; \Gamma; f]{C}{A}
        }\and
        \inferrule[T-Constr]{
            C \in \calC_A \\
            \sig(C) = (A_1, \dots, A_k)\\
            \forall i \leq k.\; \judge[\Sigma; \Gamma; f]{e_i}{\later^{n+1} A_i}
        }{
            \judge[\Sigma; \Gamma; f]{C(\overline{e})}{\later^{n} A}
        }\and
        \inferrule[T-Let]{
            \judge[\Sigma; \Gamma; f]{e}{\tau}\\
            \judge[\Sigma; \Gamma, x : \tau; f]{e'}{\tau'}
        }{
            \judge[\Sigma; \Gamma; f]{\letin{x}{e}{e'}}{\tau'}
        }\and
        \inferrule[T-Call]{
            j \geq 1 \text{ if } g = f \text{ else } j \geq 0\\
            \Sigma(g) = (\later^{m_1} A_1, \dots, \later^{m_k} A_k) \to \later^n A'\\
            \forall i \leq k.\; \judge[\Sigma; \Gamma; f]{e_i}{\later^{m_i + j} A_i}
        }{
            \judge[\Sigma; \Gamma; f]{g(\overline{e})}{\later^{n + j} A'}
        }\and
        \inferrule[T-Case]{
            \judge[\Sigma; \Gamma; f]{e}{\later^n A}\\
            \calC_A = \{C_1, \dots, C_k\}\\
            \forall i \leq k.\; \sig(C_i) = (A_{i1}, \dots, A_{ik_i}) \\
            n \leq m\\
            \forall i \leq k.\; \judge[\Sigma; \Gamma, x_{i1} : \later^{n+1} A_{i1}, \dots, x_{ik_i} : \later^{n+1} A_{ik_i}; f]{e_i}{\later^m A'}
        }{
            \judge[\Sigma; \Gamma; f]{\case{e}{C_i}{x_i}{e_i}^k}{\later^m A'}
        }\and
        \inferrule[T-Prog-Main]{
            \judge[\Sigma; \varnothing; \cdot]{e}{\tau}
        }{
            \judge[\Sigma]{e}{\tau}
        }\and
        \inferrule[T-Defn]{
            f \not\in \dom(\Sigma)\\
            \Sigma' = \Sigma, f : \overline{\tau} \to \tau'\\
            \judge[\Sigma';\; \overline{x : \tau};\; f]{e}{\tau'}\\
            \judge[\Sigma']{P}{\tau''}
        }{
            \judge[\Sigma]{\defin f(\overline{x : \tau}) : \tau' = e;\, P}{\tau''}
        }
    \end{mathpar}
    \caption{Typing rules for the \tali language.}
    \label{fig:fullTypeRules}
\end{figure}

The complete set of typing rules for \tali is shown in \autoref{fig:fullTypeRules}.
The additional rules not found in \autoref{fig:typeRules} are
\textsc{T-Var},
\textsc{T-Flip},
\textsc{T-Geom},
\textsc{T-Constr-Null},
\textsc{T-Let},
\textsc{T-Prog-Main}, and
\textsc{T-Defn}.

An important detail of the type system is that function arguments and return values may have arbitrary modal indices in function signatures.
This is necessary because arguments may be used freely in the function body: they may be destructured in $\kw{case}$ expressions, wrapped in constructors, or passed to downstream (possibly recursive) function calls.
Without tracking the modalities of arguments and return values, a value at a higher modality could therefore be laundered into a lower one.

Modalities on function argument and return types are not polymorphic.
Since \tali is first-order, a simple shift lemma suffices to show that the soundness invariant is monotone in the modal index.

\begin{lemma}[Shift lemma]
    \label{lem:shift}
    Suppose $\judge[\Sigma; \Gamma; f]{e}{\later^n A}$ and $j \geq 0$.
    Then $\judge[\Sigma; \later^j \Gamma; f]{e}{\later^{n+j} A}$, where $\later^j \Gamma$ is $\Gamma$ where each variable in $\Gamma$ has had its index incremented by $j$.
\end{lemma}

The proof is by induction on the typing derivation.

\section{Experimental setup}

\subsection{Synthesis DSL grammars}
\label{apx:synthDSL}

Here we give the family of \emph{covering skeletons} from which the initial program search of \autoref{subsec:initialProgSearch} draws its starting programs.
Let $f$ be the function whose body is the hole, let $T$ be its return type, and let $C_1, \dots, C_m$ be the constructors of $T$ that occur in the training observations, listed in declaration order.
\autoref{fig:skeletonGrammar} gives the grammar, and \autoref{fig:cascadeExample} shows two skeletons it generates.
Each $\pi$ in the grammar denotes a fresh symbolic parameter, so distinct probabilistic choices in a skeleton have distinct parameters.

\begin{figure}[t]
    \centering
    \[
        \begin{array}{l@{\quad}r@{\quad}c@{\quad}l}
        \text{Bodies:}         & B                  &\Coloneqq& K(C_1, \dots, C_m) \mid \case{x}{D}{y}{K_D}\\
        \text{Cascades:}       & K(C_1)             &\Coloneqq& A(C_1)\\
                               & K(C_1, \dots, C_m) &         & \ite{\flip \pi}{A(C_1)}{K(C_2, \dots, C_m)}\\
        \text{Arms:}           & A(C)               &\Coloneqq& \letin{\overline{w}}{\overline{I}}{C(\overline{F})}\\
        \text{Fields:}         & F                  &\Coloneqq& f(\overline{a}) \mid w \mid I \mid \flip \pi \mid g(\overline{a}) \mid L(L_1, \dots, L_r)\\
        \text{Labels:}         & L(L_1)             &\Coloneqq& L_1\\
                               & L(L_1, \dots, L_r) &         & \ite{\flip \pi}{L_1}{L(L_2, \dots, L_r)}\\
        \text{Integer forms:}  & I                  &\Coloneqq& \geom \pi \mid \fn{min}(\geom \pi, h) \mid \fn{min}(l + \geom \pi, h)\\
                               &                    &         & \mid s + \geom \pi \mid \fn{min}(s + \geom \pi, h) \\
                               &                    &         & \mid \ite{\flip \pi}{c}{\geom \pi'}\\
        \text{Call arguments:} & a                  &\Coloneqq& y \mid t \mid x \mid x + 1 \mid x - 1
        \end{array}
    \]
    \caption{%
        Grammar of covering skeletons for a hole of return type $T$ in function $f$.
        Productions for fields $F$ and arguments $a$ are selected by type:
        a field of type $T$ is a self-call $f(\overline{a})$;
        an $\mathsf{Int}$ field is an integer form $I$, or a $\kw{let}$-bound variable $w$ holding one;
        a $\mathsf{Bool}$ field is $\flip \pi$;
        for a field of any other type $S$, the grammar uses a call $g(\overline{a})$ to a generator $g$ for $S$ when one is in scope. Otherwise, if $S$ is an enumeration $L_1 \mid \dots \mid L_r$, it uses a label cascade $L(L_1, \dots, L_r)$.
        We write $\fn{min}(e, h)$ for $\letin{v}{e}{\ite{v \leq h}{v}{h}}$.
    }
    \label{fig:skeletonGrammar}
\end{figure}

\paragraph{Cascades}
The body of a basic skeleton is a \emph{cascade}: a right-nested chain of $\kw{if}$ expressions with one arm for each observed constructor of $T$.
Each guard is a bare $\flip \pi$, never a test on the function's arguments. A guard based on the arguments could make an arm unreachable and remove its constructor from the support.
A cascade over $m$ constructors therefore makes $m - 1$ choices and gives every observed constructor nonzero probability at every call.
Together with self-calls in every field of type $T$, this makes the skeleton covering: every observation built from the observed constructors has nonzero likelihood.

If $f$ has an argument $x$ of algebraic type, the body instead destructures $x$ and places a separate cascade $K_D$ in each arm $D(\overline{y})$.
When $x$ is inductive, each self-call passes a destructured field $y$ of the same type, so recursion descends.
An arm that binds no such field is a base case, and its cascade contains only observed constructors without fields of type $T$.
When $x$ is not inductive (for example, the tag introduced by call-site splitting in \autoref{subsec:initialProgSearch}), every arm retains all observed constructors, and each tag constructor receives its own independently parameterized cascade.

\paragraph{Fields}
Non-integer fields are filled as described in \autoref{fig:skeletonGrammar}.
Label cascades for enumerations range over \emph{all} constructors, not only those observed in training, just as $\geom \pi$ covers all of $\N$ rather than only the observed integers. Thus, a label that appears only in held-out data still receives nonzero probability, and parameter fitting removes the mass assigned to unused labels.

Each integer field takes one of the forms $I$.
In these forms, $h$ and $l$ range over integer variables in scope that were not drawn in the current arm (typically the integer parameters of $f$), $s$ ranges over integer fields of the same constructor that were bound earlier in the arm, and $c$ is the most common field value in the training set.
The forms $s + \geom \pi$ and $\fn{min}(s + \geom \pi, h)$ let one field depend on another. For example, the ordered pair $\letin{w_1}{\geom \pi_1}{\con{Pair}(w_1, w_1 + \geom \pi_2)}$ is generated this way.
The literal-or-draw form $\ite{\flip \pi}{c}{\geom \pi'}$ mixes a point mass at $c$ with a geometric draw. It fits fields whose values concentrate at a mode far from zero more compactly than a geometric draw alone, while preserving the field's full support.
Integer fields are $\kw{let}$-bound so that later fields and call arguments may refer to them; a binding referenced only by its own field is inlined.

\paragraph{Call arguments}
An argument of the inductive type being destructured is the corresponding field $y$, and a tag argument is the tag $t$ determined by the call site.
An integer argument is $x$, $x + 1$, or $x - 1$, where $x$ is an integer variable in scope, including a field bound in the current arm.
This is what allows a drawn value to be threaded into the recursion.

\paragraph{Search space}
The \emph{default} skeleton chooses $\geom \pi$ for every integer field and passes each integer parameter of $f$ through unchanged at every self-call.
The initial program search enumerates (or performs coordinate descent over) the product of the integer forms and integer arguments at each choice point.
Because the options at each choice point depend only on the skeleton, not on choices made elsewhere, this space is fixed before the search begins.

\begin{figure}[t]
    \centering
    \[
        \begin{array}{ll}
            \text{(a)} & \defin \fn{genExpr}(h : \mathsf{Int}) : \text{Expr} = \\
            & \quad \kw{if} \flip \pi_1\;\kw{then} \con{Lit}(\fn{min}(\geom \pi_2, h))\\
            & \quad \kw{else} \kw{if} \flip \pi_3\;\kw{then} \con{Var}(\ite{\flip \pi_4}{\con{X}}{\ite{\flip \pi_5}{\con{Y}}{\con{Z}}})\\
            & \quad \kw{else} \con{Add}(\fn{genExpr}(h), \fn{genExpr}(h))\\[1.5ex]
            \text{(b)} & \defin \fn{sorted}(l : \mathsf{Int}, h : \mathsf{Int}) : \text{List} = \\
            & \quad \kw{if} \flip \pi_1\;\kw{then} \con{Nil}\\
            & \quad \kw{else} \letin{x}{\fn{min}(l + \geom \pi_2, h)}{\con{Cons}(x, \fn{sorted}(x, h))}
        \end{array}
    \]
    \caption{%
        Example covering skeletons.
        (a) A cascade over the observed constructors $\con{Lit}$, $\con{Var}$, and $\con{Add}$ of $\text{Expr}$.
        The $\con{Var}$ field has the enumeration type $\con{X} \mid \con{Y} \mid \con{Z}$ and is filled by a label cascade, and both $\con{Add}$ fields are self-calls.
        (b) The sorted-list skeleton of \autoref{subsec:initialProgSearch}.
        It differs from the default skeleton $\ite{\flip \pi_1}{\con{Nil}}{\con{Cons}(\geom \pi_2, \fn{sorted}(l, h))}$ at two choice points: the integer form of the head, and the first argument of the self-call, which threads the bound variable $x$.
    }
    \label{fig:cascadeExample}
\end{figure}

\subsection{Transformer setup}
\label{apx:transformerSetup}

\subsubsection{Tokenization}
\label{apx:tokenization}

Training samples in our datasets are labeled trees.
Because constructors have known arities, labeled trees correspond bijectively to sequences of labels in preorder.
We serialize each tree with a preorder traversal and assign each label a unique token in the transformer's vocabulary.
For trees with integer labels, we instantiate a token for each integer between the minimum and maximum labels in the dataset.

\subsubsection{Pretraining}
\label{apx:pretraining}

Each pretrained transformer is pretrained on a corpus of 10,000 samples from the corpus generator under test (i.e. program synthesis or random ADT).
The program synthesis pretraining corpus generator is trained on $n$ samples, varied during testing.
Following pretraining, the transformer is fine-tuned on the original $n$ training samples regardless of pretraining method.

\subsubsection{Hyperparameters}
\label{apx:hyperparameters}

All transformer models in this paper use the PyTorch \texttt{nn.Transformer} implementation.
Each transformer has two hidden layers of size 256, eight attention heads, and a maximum context window of 1,000 tokens.
We train the pretrained transformers using Adam~\citep{kingmaAdam2017} with a batch size of 32 and a constant learning rate of $10^{-3}$. We use five pretraining epochs and up to 20 fine-tuning epochs, with early stopping on a validation set (patience 3, minimum delta $10^{-4}$).

The full-pool transformers have the same architecture but use a cosine-decay schedule with 1,000 linear warmup steps.
We clip gradients at an $L^2$ norm of $1$ and use a dropout rate of $0.1$ during training.
The full-pool transformers are trained for up to 30 epochs using the same early stopping setup.

\section{Experimental results}
\label{apx:experimentData}

This appendix gives the numbers behind the figures of \autoref{sec:evaluation}.
Real-data tables report the nine real-world cells: SPARQL, and the regex dataset in two character encodings crossed with four training subsets.
The \emph{exact} encoding keeps every literal character, while \emph{kind} keeps only the class of letters and digits; the \emph{random} subset is a uniform sample of the corpus, and \emph{repeat}, \emph{range} and \emph{backref} are drawn from regexes containing bounded repeats, set ranges and backreferences, respectively.
As in \autoref{subsec:baseline-description}, NLLs are in nats per tree, excesses and differences are medians across seeds, and rates and recalls are means across seeds.

\subsection{Synthesized program fit}

\autoref{tbl:apxFit} gives the data behind \autoref{fig:fitRecall}: for each real-world cell and training-set size, the program's paired excess NLL over the matched-$n$ untrained transformer, beside the program's support recall.
Because a median excess can hide a heavy tail of badly fit trees, the last column also reports CVaR$_{10}$, the mean paired excess over the worst 10\% of covered test trees, at $n=300$.
\autoref{tbl:apxFitReference} gives the absolute NLLs at $n=300$, alongside the reference models.

\begin{table}[h]
  \centering
  \scriptsize
  \setlength{\tabcolsep}{3pt}
  \begin{tabular*}{\linewidth}{@{\extracolsep{\fill}}lrrrrrrrrr}
\toprule
 & \multicolumn{4}{c}{paired excess, by $n$} & \multicolumn{4}{c}{recall, by $n$} & CVaR$_{10}$ \\
\cmidrule(lr){2-5}\cmidrule(lr){6-9}\cmidrule(lr){10-10}
Cell & 10 & 30 & 100 & 300 & 10 & 30 & 100 & 300 & 300 \\
\midrule
SPARQL & $-10.95$ & $\mathbf{-23.07}$ & $\mathbf{-20.94}$ & $\mathbf{-6.05}$ & $.623$ & $.864$ & $.974$ & $.997$ & $22.3$ \\
exact/repeat & $2.38$ & $\mathbf{-6.34}$ & $\mathbf{-6.57}$ & $\mathbf{-1.13}$ & $.635$ & $.902$ & $.981$ & $.998$ & $23.9$ \\
exact/range & $11.21$ & $-2.43$ & $\mathbf{-3.44}$ & $0.22$ & $.648$ & $.927$ & $.983$ & $.998$ & $26.7$ \\
exact/backref & $21.85$ & $5.87$ & $-0.80$ & $5.02$ & $.700$ & $.930$ & $.997$ & $1.000$ & $31.3$ \\
exact/random & $5.04$ & $1.81$ & $0.30$ & $2.22$ & $.600$ & $.896$ & $.985$ & $.998$ & $21.5$ \\
kind/repeat & $-2.90$ & $\mathbf{-6.21}$ & $\mathbf{-5.01}$ & $-0.58$ & $.624$ & $.901$ & $.979$ & $.999$ & $22.8$ \\
kind/range & $7.36$ & $-3.31$ & $\mathbf{-3.85}$ & $-0.24$ & $.703$ & $.912$ & $.983$ & $.998$ & $22.9$ \\
kind/backref & $19.83$ & $6.43$ & $-0.39$ & $4.60$ & $.689$ & $.915$ & $.992$ & $1.000$ & $30.1$ \\
kind/random & $5.50$ & $4.01$ & $-1.47$ & $0.46$ & $.601$ & $.888$ & $.990$ & $.998$ & $17.9$ \\
\bottomrule
\end{tabular*}

  \caption{%
    Fit of synthesized programs on real data.
    \emph{Paired excess:} median, across seeds, of the program's mean excess NLL over the untrained transformer trained on the same $n$ trees, on the test trees the program covers (negative favors the program); bold where the program wins with sign-test $p < 0.01$.
    \emph{Recall:} mean fraction of test trees to which the program assigns nonzero probability.
    \emph{CVaR$_{10}$:} median, across seeds, of the mean excess over the worst 10\% of covered test trees, at $n=300$.
  }
  \label{tbl:apxFit}
\end{table}

\begin{table}[h]
  \centering
  \scriptsize
  \begin{tabular*}{\linewidth}{@{\extracolsep{\fill}}lrrr}
\toprule
\multicolumn{4}{l}{\emph{Covered trees}} \\
Cell & Program & No pretraining &  \\
\cmidrule(lr){1-1}\cmidrule(lr){2-3}
SPARQL & $73.9$ & $79.7$ &  \\
exact/repeat & $70.7$ & $72.2$ &  \\
exact/range & $66.2$ & $66.2$ &  \\
exact/backref & $87.0$ & $81.7$ &  \\
exact/random & $67.1$ & $64.6$ &  \\
kind/repeat & $62.2$ & $62.8$ &  \\
kind/range & $57.0$ & $57.1$ &  \\
kind/backref & $74.6$ & $70.5$ &  \\
kind/random & $49.1$ & $48.5$ &  \\
\midrule
\multicolumn{4}{l}{\emph{All trees, matched $n$}} \\
Cell & No pretraining & Random-pretrained & Program-pretrained \\
\cmidrule(lr){1-1}\cmidrule(lr){2-4}
SPARQL & $79.7$ & $70.3$ & $56.3$ \\
exact/repeat & $72.2$ & $74.3$ & $60.1$ \\
exact/range & $66.2$ & $67.6$ & $57.1$ \\
exact/backref & $81.7$ & $82.6$ & $74.0$ \\
exact/random & $64.7$ & $64.6$ & $59.9$ \\
kind/repeat & $62.9$ & $65.0$ & $52.0$ \\
kind/range & $57.1$ & $57.5$ & $48.6$ \\
kind/backref & $70.5$ & $71.4$ & $62.2$ \\
kind/random & $48.5$ & $48.5$ & $42.7$ \\
\midrule
\multicolumn{4}{l}{\emph{All trees, references}} \\
Cell & PTA (smoothed) & Full pool & Full pool (2-layer) \\
\cmidrule(lr){1-1}\cmidrule(lr){2-4}
SPARQL & $175.4$ & $23.3$ & $26.2$ \\
exact/repeat & $138.2$ & -- & -- \\
exact/range & $134.6$ & -- & -- \\
exact/backref & $156.3$ & -- & -- \\
exact/random & $115.6$ & $29.5$ & $30.8$ \\
kind/repeat & $126.9$ & -- & -- \\
kind/range & $123.5$ & -- & -- \\
kind/backref & $140.6$ & -- & -- \\
kind/random & $96.1$ & $24.2$ & $25.0$ \\
\bottomrule
\end{tabular*}

  \caption{%
    Median test NLL (nats per tree) on real data at $n=300$.
    \emph{Covered trees:} the synthesized program and the untrained transformer, both on the test trees the program covers.
    \emph{All trees, matched $n$:} the three transformers trained on the same 300 trees.
    \emph{All trees, references:} a smoothed probabilistic tree automaton (PTA) fit to the same 300 trees, and the full-pool transformer in its six-layer (used in the figures) and two-layer configurations, which were trained only for SPARQL and the \emph{random} regex cells.
  }
  \label{tbl:apxFitReference}
\end{table}

\autoref{tbl:apxAblation} ablates the synthesis procedure of \autoref{sec:synthesis} on real data.
It builds up from the initial program search alone (\emph{seed}), adding stochastic search (\emph{chain}) and call-site parent annotation (\emph{annot.}), to the full method, which adds the validation-set selection step (\emph{guard}).
Parent annotation is the largest single effect, and its sign depends on $n$: at $n=300$ it removes 16--28 nats of excess on every cell, but at $n=10$ it costs fit on every cell.
The chain recovers a few nats of that cost at $n=10$ and has little effect at $n=300$; the guard changes almost nothing on real data.

\begin{table}[h]
  \centering
  \scriptsize
  \setlength{\tabcolsep}{3pt}
  \begin{tabular*}{\linewidth}{@{\extracolsep{\fill}}lrrrrr}
\toprule
 &  & seed & seed & seed+annot. & seed+annot. \\
Cell & seed & +chain & +annot. & +chain & +chain+guard \\
\midrule
\multicolumn{6}{l}{\emph{$n=10$}} \\
SPARQL & $-13.7$\,{\scriptsize (.65)} & $-13.7$\,{\scriptsize (.64)} & $-7.2$\,{\scriptsize (.65)} & $-11.0$\,{\scriptsize (.62)} & $-11.0$\,{\scriptsize (.62)} \\
exact/repeat & $-19.1$\,{\scriptsize (.66)} & $-16.6$\,{\scriptsize (.62)} & $6.6$\,{\scriptsize (.66)} & $2.4$\,{\scriptsize (.63)} & $2.4$\,{\scriptsize (.63)} \\
exact/range & $2.1$\,{\scriptsize (.68)} & $1.7$\,{\scriptsize (.66)} & $15.9$\,{\scriptsize (.68)} & $11.2$\,{\scriptsize (.65)} & $11.2$\,{\scriptsize (.65)} \\
exact/backref & $-3.8$\,{\scriptsize (.73)} & $-2.7$\,{\scriptsize (.66)} & $24.9$\,{\scriptsize (.73)} & $21.9$\,{\scriptsize (.70)} & $21.9$\,{\scriptsize (.70)} \\
exact/random & $-0.8$\,{\scriptsize (.63)} & $0.0$\,{\scriptsize (.61)} & $6.7$\,{\scriptsize (.63)} & $5.0$\,{\scriptsize (.60)} & $5.0$\,{\scriptsize (.60)} \\
kind/repeat & $-14.7$\,{\scriptsize (.65)} & $-13.5$\,{\scriptsize (.61)} & $-0.6$\,{\scriptsize (.65)} & $-2.9$\,{\scriptsize (.62)} & $-2.9$\,{\scriptsize (.62)} \\
kind/range & $-1.0$\,{\scriptsize (.73)} & $-1.1$\,{\scriptsize (.68)} & $8.6$\,{\scriptsize (.73)} & $7.4$\,{\scriptsize (.70)} & $7.4$\,{\scriptsize (.70)} \\
kind/backref & $-4.4$\,{\scriptsize (.71)} & $-1.0$\,{\scriptsize (.66)} & $22.8$\,{\scriptsize (.71)} & $19.8$\,{\scriptsize (.69)} & $19.8$\,{\scriptsize (.69)} \\
kind/random & $1.1$\,{\scriptsize (.62)} & $1.6$\,{\scriptsize (.60)} & $5.4$\,{\scriptsize (.62)} & $5.5$\,{\scriptsize (.60)} & $5.5$\,{\scriptsize (.60)} \\
\midrule
\multicolumn{6}{l}{\emph{$n=300$}} \\
SPARQL & $22.0$\,{\scriptsize (1)} & $21.5$\,{\scriptsize (1)} & $-5.9$\,{\scriptsize (1)} & $-6.1$\,{\scriptsize (1)} & $-6.0$\,{\scriptsize (1)} \\
exact/repeat & $16.3$\,{\scriptsize (1)} & $16.4$\,{\scriptsize (1)} & $-0.4$\,{\scriptsize (1)} & $-1.1$\,{\scriptsize (1)} & $-1.1$\,{\scriptsize (1)} \\
exact/range & $19.8$\,{\scriptsize (1)} & $19.8$\,{\scriptsize (1)} & $0.3$\,{\scriptsize (1)} & $0.2$\,{\scriptsize (1)} & $0.2$\,{\scriptsize (1)} \\
exact/backref & $27.1$\,{\scriptsize (1)} & $26.9$\,{\scriptsize (1)} & $5.0$\,{\scriptsize (1)} & $5.0$\,{\scriptsize (1)} & $5.0$\,{\scriptsize (1)} \\
exact/random & $18.5$\,{\scriptsize (1)} & $18.5$\,{\scriptsize (1)} & $2.3$\,{\scriptsize (1)} & $2.2$\,{\scriptsize (1)} & $2.2$\,{\scriptsize (1)} \\
kind/repeat & $16.1$\,{\scriptsize (1)} & $15.8$\,{\scriptsize (1)} & $-0.1$\,{\scriptsize (1)} & $-0.6$\,{\scriptsize (1)} & $-0.6$\,{\scriptsize (1)} \\
kind/range & $19.7$\,{\scriptsize (1)} & $19.7$\,{\scriptsize (1)} & $0.0$\,{\scriptsize (1)} & $-0.2$\,{\scriptsize (1)} & $-0.2$\,{\scriptsize (1)} \\
kind/backref & $26.6$\,{\scriptsize (1)} & $26.5$\,{\scriptsize (1)} & $4.7$\,{\scriptsize (1)} & $4.6$\,{\scriptsize (1)} & $4.6$\,{\scriptsize (1)} \\
kind/random & $18.4$\,{\scriptsize (1)} & $18.4$\,{\scriptsize (1)} & $0.5$\,{\scriptsize (1)} & $0.5$\,{\scriptsize (1)} & $0.5$\,{\scriptsize (1)} \\
\bottomrule
\end{tabular*}

  \caption{%
    Ablation of the synthesis procedure at $n=10$ and $n=300$: median paired excess NLL over the untrained transformer on the test trees each program covers (negative favors the program), with mean support recall in parentheses.
    \emph{seed}: initial program search only; \emph{+chain}: followed by stochastic search; \emph{+annot.}: with call-site parent annotation; \emph{+guard}: with validation-set selection among the top-$k$ programs.
    The rightmost column is the method evaluated in \autoref{sec:evaluation}.
  }
  \label{tbl:apxAblation}
\end{table}

\subsection{Structure recovery}

\autoref{tbl:apxGridNovel} breaks the grid's novel-sample violation rates in \autoref{tbl:grid} down by cell: each entry gives the range of per-cell rates within a block, rather than the block mean.
It also adds the seed-only program, which stops after the initial program search.
The seed-only program recovers the V2 invariants exactly on some cells at $n=300$, but nearly every novel sample it produces on V1 violates the invariant.

\begin{table}[h]
  \centering
  \scriptsize
  \setlength{\tabcolsep}{3pt}
  \begin{tabular*}{\linewidth}{@{\extracolsep{\fill}}lcccccccc}
\toprule
 & \multicolumn{2}{c}{No pretraining} & \multicolumn{2}{c}{Random-pretrained} & \multicolumn{2}{c}{Program} & \multicolumn{2}{c}{Seed only} \\
\cmidrule(lr){2-3}\cmidrule(lr){4-5}\cmidrule(lr){6-7}\cmidrule(lr){8-9}
Block & $n=10$ & $n=300$ & $n=10$ & $n=300$ & $n=10$ & $n=300$ & $n=10$ & $n=300$ \\
\midrule
S1--S3 & .22--.41 & .01--.13 & .00--.07 & .00--.06 & .02--.06 & .00--.04 & .01--.06 & .00--.10 \\
S4 & .37--.49 & .09--.28 & .14--.40 & .09--.45 & .12--.28 & .07--.39 & .17--.28 & .28--.42 \\
V1 & .39--.76 & .26--.35 & .47--.73 & .26--.36 & .38--.73 & .82--.90 & .46--.78 & .98--.99 \\
V2 & .33--.65 & .18--.46 & .32--.39 & .21--.54 & .12--.18 & .13--.30 & .13--.14 & .00--.30 \\
V3 & .33--.99 & .11--.87 & .21--.98 & .64--.99 & .29--.99 & .71--1.00 & .27--.99 & .72--1.00 \\
\bottomrule
\end{tabular*}

  \caption{%
    Mean rate at which novel, well-formed samples fall outside the ground truth's support on the synthetic grid, per block, arm and training-set size (lower is better).
    Where a block has several cells, the entry gives the range of the per-cell means.
    \emph{Program} is the synthesized program, as in \autoref{tbl:grid}; \emph{Seed only} is the program produced by the initial program search alone.
  }
  \label{tbl:apxGridNovel}
\end{table}

\autoref{tbl:apxValidity} reports how often each model's samples are valid in the real-world domains, at the smallest and largest $n$.
Besides the method, it includes two synthesis variants: the seed search with parent annotation but without stochastic search, and the full search without parent annotation.
The unannotated program compiles slightly more often as a regex, but rarely produces a valid SPARQL query, since the root query pattern is then drawn from the same rule as every nested one.
The last column counts transformer samples that do not decode to a tree at all; programs produce well-typed trees by construction.

\begin{table}[h]
  \centering
  \scriptsize
  \setlength{\tabcolsep}{3pt}
  \begin{tabular*}{\linewidth}{@{\extracolsep{\fill}}lrrrrrrrr}
\toprule
 & \multicolumn{2}{c}{Regex} & \multicolumn{4}{c}{SPARQL} & \multicolumn{2}{c}{} \\
\cmidrule(lr){2-3}\cmidrule(lr){4-7}
 & \multicolumn{2}{c}{compiles} & \multicolumn{2}{c}{valid} & \multicolumn{2}{c}{well-formed} & \multicolumn{2}{c}{malformed} \\
\cmidrule(lr){2-3}\cmidrule(lr){4-5}\cmidrule(lr){6-7}\cmidrule(lr){8-9}
Arm & $n=10$ & $n=300$ & $n=10$ & $n=300$ & $n=10$ & $n=300$ & $n=10$ & $n=300$ \\
\midrule
Program & $.647$ & $.609$ & $.480$ & $.410$ & $.608$ & $.564$ & -- & -- \\
Seed + annot./lit. & $.647$ & $.612$ & $.466$ & $.421$ & $.592$ & $.584$ & -- & -- \\
Unannotated & $.688$ & $.693$ & $.028$ & $.016$ & $.034$ & $.018$ & -- & -- \\
Program-pretrained & $.476$ & $.460$ & $.355$ & $.450$ & $.431$ & $.545$ & $.392$ & $.344$ \\
Random-pretrained & $.205$ & $.121$ & $.027$ & $.211$ & $.030$ & $.252$ & $.763$ & $.633$ \\
No pretraining & $.003$ & $.042$ & $.000$ & $.022$ & $.000$ & $.023$ & $.978$ & $.900$ \\
Full pool & \multicolumn{2}{c}{$.996$} & \multicolumn{2}{c}{$.966$} & \multicolumn{2}{c}{$.989$} & \multicolumn{2}{c}{$.009$} \\
Full pool (2-layer) & \multicolumn{2}{c}{$.986$} & \multicolumn{2}{c}{$.943$} & \multicolumn{2}{c}{$.976$} & \multicolumn{2}{c}{$.021$} \\
\midrule
Pool & \multicolumn{2}{c}{--} & \multicolumn{2}{c}{$.981$} & \multicolumn{2}{c}{$1.000$} & \multicolumn{2}{c}{--} \\
\bottomrule
\end{tabular*}

  \caption{%
    Fraction of samples that are valid in the real-world domains, at $n=10$ and $n=300$ (mean across seeds).
    \emph{Regex compiles:} accepted by Python's \texttt{re.compile}.
    \emph{SPARQL valid:} well-formed and satisfying the four scoping rules of \autoref{fig:structureRules}; \emph{well-formed:} printable as a query; \emph{malformed:} a transformer sample whose token sequence does not decode to a tree.
    \emph{Program}, \emph{Seed + annot./lit.} and \emph{Unannotated} are synthesized programs; the full-pool transformers are trained once, so a single entry spans both $n$.
    \emph{Pool:} the SPARQL training data itself.
  }
  \label{tbl:apxValidity}
\end{table}

\autoref{tbl:apxRules} gives the per-rule violation rates plotted in \autoref{fig:structureRules}.

\begin{table}[h]
  \centering
  \scriptsize
  \begin{tabular*}{\linewidth}{@{\extracolsep{\fill}}lrrrrrr}
\toprule
\emph{Regex} & \multicolumn{3}{c}{local} & \multicolumn{1}{c}{agreement} \\
\cmidrule(lr){2-4}\cmidrule(lr){5-5}
Arm & no compile & repeat & range & backref. &  &  \\
\cmidrule(lr){1-1}\cmidrule(lr){2-5}
Program & $.494$ & $.005$ & $.005$ & $.767$ &  &  \\
Program-pretrained & $.500$ & $.025$ & $.002$ & $.246$ &  &  \\
Random-pretrained & $.792$ & $.307$ & $.006$ & $.388$ &  &  \\
No pretraining & $.949$ & $.093$ & $.000$ & $.288$ &  &  \\
Full pool & $.005$ & $.002$ & $.002$ & $.062$ &  &  \\
\midrule
\emph{SPARQL} & \multicolumn{2}{c}{local} & \multicolumn{4}{c}{agreement} \\
\cmidrule(lr){2-3}\cmidrule(lr){4-7}
Arm & W & S1 & S2 & S3 & S4 & N1 \\
\cmidrule(lr){1-1}\cmidrule(lr){2-7}
Program & $.436$ & $.060$ & $.893$ & $.309$ & $.652$ & $.986$ \\
Program-pretrained & $.455$ & $.030$ & $.897$ & $.128$ & $.626$ & $.550$ \\
Random-pretrained & $.748$ & $.082$ & $.919$ & $.087$ & $.509$ & $.482$ \\
No pretraining & $.977$ & $.000$ & $.806$ & $.000$ & $.260$ & $.184$ \\
Full pool & $.011$ & $.004$ & $.030$ & $.024$ & $.121$ & $.005$ \\
Pool & $.000$ & $.007$ & $.005$ & $.000$ & $.106$ & $.000$ \\
\bottomrule
\end{tabular*}

  \caption{%
    Fraction of samples violating each validity rule at $n=300$ (mean across seeds; lower is better), as in \autoref{fig:structureRules}.
    \emph{Regex:} fails \texttt{re.compile}; repeat \texttt{\{m,n\}} with $m > n$; set range with reversed bounds; backreference to an unclosed group.
    \emph{SPARQL:} not printable as a query (W); misplaced or nested aggregate (S1); aggregate query projects a non-\texttt{GROUP BY} variable (S2); \texttt{BIND} to an in-scope variable (S3); \texttt{AS ?v} rebinds a bound variable (S4); variables not numbered in first-use order (N1).
    Except for compilation and W, rates are among samples containing the rule's construct.
    \emph{Pool:} the SPARQL training data, which includes some queries the specification forbids.
  }
  \label{tbl:apxRules}
\end{table}

\subsection{Data augmentation}

\autoref{tbl:apxUse} gives the data behind \autoref{fig:useLearnCurve}.
The first two blocks compare the program-pretrained transformer with each control trained on the same $n$ trees, and the last gives its own test NLL beside the full-pool reference.
Every transformer covers every test tree, so these comparisons are over all test trees and carry no recall caveat.

\begin{table}[h]
  \centering
  \scriptsize
  \setlength{\tabcolsep}{4pt}
  \begin{tabular*}{\linewidth}{@{\extracolsep{\fill}}lrrrrr}
\toprule
Cell & $n=10$ & $n=30$ & $n=100$ & $n=300$ &  \\
\midrule
\multicolumn{5}{l}{\emph{$\Delta$ NLL vs.\ random-pretrained (fraction of seeds won)}} &  \\
SPARQL & $-19.9$\,{\scriptsize .93} & $-28.7$\,{\scriptsize 1} & $-20.7$\,{\scriptsize 1} & $-14.3$\,{\scriptsize 1} &  \\
exact/repeat & $-18.3$\,{\scriptsize 1} & $-22.8$\,{\scriptsize 1} & $-17.2$\,{\scriptsize 1} & $-14.2$\,{\scriptsize 1} &  \\
exact/range & $-12.0$\,{\scriptsize .96} & $-17.6$\,{\scriptsize 1} & $-12.2$\,{\scriptsize 1} & $-10.3$\,{\scriptsize 1} &  \\
exact/backref & $-10.1$\,{\scriptsize .89} & $-14.0$\,{\scriptsize 1} & $-9.3$\,{\scriptsize 1} & $-8.8$\,{\scriptsize 1} &  \\
exact/random & $2.7$\,{\scriptsize .44} & $-8.6$\,{\scriptsize .96} & $-5.7$\,{\scriptsize 1} & $-5.1$\,{\scriptsize 1} &  \\
kind/repeat & $-17.6$\,{\scriptsize .93} & $-21.9$\,{\scriptsize 1} & $-15.4$\,{\scriptsize 1} & $-13.3$\,{\scriptsize 1} &  \\
kind/range & $-14.4$\,{\scriptsize .96} & $-16.7$\,{\scriptsize 1} & $-11.5$\,{\scriptsize 1} & $-9.4$\,{\scriptsize 1} &  \\
kind/backref & $-8.7$\,{\scriptsize .85} & $-14.0$\,{\scriptsize 1} & $-9.9$\,{\scriptsize 1} & $-9.4$\,{\scriptsize 1} &  \\
kind/random & $-1.6$\,{\scriptsize .56} & $-9.2$\,{\scriptsize 1} & $-6.3$\,{\scriptsize 1} & $-6.0$\,{\scriptsize 1} &  \\
\midrule
\multicolumn{5}{l}{\emph{$\Delta$ NLL vs.\ no pretraining (fraction of seeds won)}} &  \\
SPARQL & $-35.4$\,{\scriptsize 1} & $-45.5$\,{\scriptsize 1} & $-38.4$\,{\scriptsize 1} & $-23.6$\,{\scriptsize 1} &  \\
exact/repeat & $-25.9$\,{\scriptsize 1} & $-21.7$\,{\scriptsize 1} & $-15.5$\,{\scriptsize 1} & $-11.8$\,{\scriptsize 1} &  \\
exact/range & $-16.4$\,{\scriptsize 1} & $-17.4$\,{\scriptsize 1} & $-12.1$\,{\scriptsize 1} & $-9.1$\,{\scriptsize 1} &  \\
exact/backref & $-13.5$\,{\scriptsize 1} & $-12.4$\,{\scriptsize 1} & $-9.2$\,{\scriptsize 1} & $-7.7$\,{\scriptsize 1} &  \\
exact/random & $-5.1$\,{\scriptsize .70} & $-11.2$\,{\scriptsize 1} & $-6.5$\,{\scriptsize 1} & $-5.1$\,{\scriptsize 1} &  \\
kind/repeat & $-22.6$\,{\scriptsize .96} & $-19.6$\,{\scriptsize 1} & $-15.1$\,{\scriptsize 1} & $-10.9$\,{\scriptsize 1} &  \\
kind/range & $-16.9$\,{\scriptsize 1} & $-16.4$\,{\scriptsize 1} & $-12.2$\,{\scriptsize 1} & $-8.7$\,{\scriptsize 1} &  \\
kind/backref & $-12.5$\,{\scriptsize .93} & $-12.0$\,{\scriptsize 1} & $-9.8$\,{\scriptsize 1} & $-8.4$\,{\scriptsize 1} &  \\
kind/random & $-6.5$\,{\scriptsize .59} & $-9.8$\,{\scriptsize 1} & $-7.0$\,{\scriptsize 1} & $-5.8$\,{\scriptsize 1} &  \\
\midrule
\multicolumn{5}{l}{\emph{Student test NLL}} & full pool \\
SPARQL & $103.2$ & $75.2$ & $63.1$ & $56.3$ & $23.3$ \\
exact/repeat & $97.5$ & $74.4$ & $66.6$ & $60.1$ & -- \\
exact/range & $86.3$ & $69.4$ & $62.3$ & $57.1$ & -- \\
exact/backref & $112.7$ & $91.8$ & $83.0$ & $74.0$ & -- \\
exact/random & $92.1$ & $71.4$ & $64.6$ & $59.9$ & $29.5$ \\
kind/repeat & $85.5$ & $65.9$ & $57.8$ & $52.0$ & -- \\
kind/range & $75.3$ & $60.0$ & $53.2$ & $48.6$ & -- \\
kind/backref & $98.7$ & $79.9$ & $70.1$ & $62.2$ & -- \\
kind/random & $74.4$ & $55.0$ & $47.4$ & $42.7$ & $24.2$ \\
\bottomrule
\end{tabular*}

  \caption{%
    Pretraining on samples from a synthesized program.
    \emph{$\Delta$ NLL:} median, across seeds, of the program-pretrained transformer's test NLL minus that of the random-pretrained or untrained transformer trained on the same $n$ trees (negative favors program pretraining), with the fraction of seeds on which program pretraining wins in small type.
    \emph{Student test NLL:} median test NLL of the program-pretrained transformer, beside the full-pool transformer, which was trained only for SPARQL and the \emph{random} regex cells.
  }
  \label{tbl:apxUse}
\end{table}

\end{document}